\documentclass[journal]{IEEEtran}
\usepackage{url} 
\usepackage{lineno}
\usepackage{soul}
\usepackage{graphicx}
\usepackage{subfig}
\usepackage{subfloat}
\usepackage{amsmath}
\usepackage{booktabs}
\usepackage{autobreak}
\usepackage{multicol}
\usepackage{multirow}
\usepackage[ruled,vlined]{algorithm2e}
\allowdisplaybreaks

\ifCLASSINFOpdf
\else
\fi

\begin{document}
\title{Microseismic source imaging using physics-informed neural networks with hard constraints}
\author{{Xinquan Huang and Tariq Alkhalifah}
\thanks{(\it {Corresponding author: Xinquan Huang})\\
Xinquan Huang and Tariq Alkhalifah are with the Physical Science and Engineering division, King Abdullah University of Science and Technology, email: xinquan.huang@kaust.edu.sa, tariq.alkhalifah@kaust.edu.sa).}}%

\maketitle

\begin{abstract}
Microseismic source imaging plays a significant role in passive seismic monitoring. 
However, such a process is prone to failure due to aliasing when dealing with sparsely measured data. 
Thus, we propose a direct microseismic imaging framework based on physics-informed neural networks (PINNs), which can generate focused source images, even with very sparse recordings. 
We use the PINNs to represent a multi-frequency wavefield and then apply inverse Fourier transform to extract the source image. 
To be more specific, we modify the representation of the frequency-domain wavefield to inherently satisfy the boundary conditions (the measured data on the surface) by means of a hard constraint, which helps to avoid the difficulty in balancing the data and PDE losses in PINNs. 
Furthermore, we propose the causality loss implementation with respect to depth to enhance the convergence of PINNs. 
The numerical experiments on the Overthrust model show that the method can admit reliable and accurate source imaging for single- or multiple- sources and even in passive monitoring settings. 
Compared with the time-reversal method, the results of the proposed method are consistent with numerical methods but less noisy.
Then, we further apply our method to hydraulic fracturing monitoring field data, and demonstrate that our method can correctly image the source with fewer artifacts.

\end{abstract}


\IEEEpeerreviewmaketitle

\section{Introduction} 
\IEEEPARstart{M}{icroseismic} events location is the essential foundation of passive seismic monitoring. 
The common way to locate seismic events is based on the travel time inversion \cite{lienert1986hypocenter}, which requires the time-consuming first arrival travel time picking. 
The process of picking arrivals becomes a challenge in low signal-to-noise data and in multiple-events cases. 
What's more, the resolution and accuracy of the source locations via this type of method might be relatively low \cite{miao2021development}. 
Directly making use of the full waveform information in the seismograms can help us avoid such issues and allow for higher resolution and accuracy in locating the microseismic events. 
Kao and Shan \cite{Kao2004} proposed an imaging approach based on summing the amplitudes of the measured data at the corresponding estimated arrival times without the need for phase picking, called the source scanning algorithm. 
It is widely used in seismology, e.g. for detailed earthquake rupture imaging \cite{Ishii2007}. 
Another type of method based on full waveforms is time-reversal imaging \cite{McMechan1982}. 
Its key idea is to backpropagate the waveform recorded on the surface until the energy is focused.
\cite{gajewski2005reverse} and \cite{larmat2006time} utilized the time reversal imaging (TRI) to reverse the waveform in the time domain for source locations.  
For this category of methods, the accuracy of the location results is highly determined by the imaging condition.  
In the past decade, many types of imaging conditions, including the maximum energy imaging condition (or the zero-lag cross-correlation imaging condition) \cite{Artman2010,Yang2019}, the interferometric imaging condition for sparsely sampled data \cite{Sava2011}, 
the deconvolution imaging condition \cite{Douma2014} and the geometric mean imaging condition \cite{Nakata2016} for higher resolution, the maximum variance imaging condition \cite{Wang2017} and the grouped seismic imaging condition \cite{lin2020source,chen2021compact} for noise-robust source location, have been proposed to enhance waveform-based location results. 
However, the accuracy for TRI is still an issue in case of sparse and irregular observations. The other type of waveform-based method is to invert the observed data for a more accurate velocity model \cite{kamei2014passive,kaderli2018self,song2019passive,wang2021direct}. 
These methods require massive repeated wavefield computation, and thus, they are costly. 
Besides, those grid-based methods rely on the discretization of the spatial domain, which is not flexible for irregular geometry and complex subsurface structures, and the memory cost and precision depend on the grid resolution.   

Thanks to recent advances in computing and algorithms, machine learning has achieved a lot of success in many fields, e.g., natural language processing, computer vision, and science. 
Many computer-vision-based methods have been adapted to microseismic source location tasks \cite{Perol2018,Kriegerowski2019,Zhang2020,Wang2021,zhang2022deep,chen20223d}. 
Most of these methods deal with source locations in a supervised manner. 
That is to say, they train a neural network on simulated data with labels (source locations) available and then minimize the loss using the labels. 
However, the generalization of these methods to field data is a challenge due to the lack of a sufficiently diverse dataset \cite{alkhalifah2022mlreal}. 
Labels for the field data are often attained using conventional methods and human intervention, and thus, they are prone to errors.
On the other hand, purely data-driven methods, which do not require labels, are an attractive alternative to the supervised approach.
One way to devise such an approach is by incorporating the physics priors or specifically training the neural networks with physics constraints. 

Based on the universal function approximation theory \cite{Hornik1989}, neural networks can be utilized to represent functions like the seismic wavefield, which is a key component of the source location process using time reversal imaging.
Thus, we represent the seismic wavefield with a function of the spatial coordinates and frequency, yielding significant flexibility of the simulation in irregular domains and the case with a complex governing equation, as well as a continuous solution with fine details  \cite{alkhalifah2021wavefield,song2021solving}. 
The governing equations for seismic modeling, are used as a loss function, to optimize the neural network using physics-informed neural networks (PINN framework \cite{Raissi2019}). 
PINNs have been used in seismic forward modeling and inversion \cite{song2021solving,bin2021pinneik,song2021wavefield,huang2022pinnup,rasht2022physics,sun2023implicit}.
By means of PINNs for source imaging, the subsurface velocity information is well embedded into the neural network, and it can image the source with irregular and sparse receivers. 
As it is a mesh-free method, the precision of the source location result theoretically depends only on the data without the need for fine-grid discretization, even in dealing with complex subsurface structures.

In this paper, we propose a novel direct microseismic source imaging method by means of PINNs, in which the source image is a snapshot of the time-domain wavefield. 
We represent the wavefield as a function of the spatial coordinates and frequency via a neural network (NN) and then train the NN with the Helmholtz equation as a loss function. 
Specifically, we incorporate the observed data on the surface (boundary condition for the Helmholtz equation) into the wavefield via the hard constraint and use the modified Helmholtz equation as the loss function. 
Besides, inspired by the fact that the information guiding the optimization of the NN parameters is coming from the recorded data on the surface, we propose to impose causality to the loss calculation along the depth axis, yielding an ideal from-surface-down reconstruction of the wavefield, which can improve the convergence speed and accuracy. 
We will first demonstrate the effectiveness of the proposed method via single- and multiple-passive sources scenarios on the Overthrust model using sparsely sampled data. 
We further test our proposed method on the Oklahoma Arkoma Basin Hydraulic Fracturing data to highlight the benefits and potential of the proposed method.

In summary, our main contributions are three-fold:
\begin{itemize}
    \item We develop a direct source imaging framework based on PINNs with hard constraints.
    \item To ensure a stable and reliable training process,  we use a reference frequency loss function incorporated in the hard constraint implementation and impose causality on the loss function with respect to depth.
    \item We evaluate the proposed method on the synthetic data for two different cases, as well as field data, and achieve reliable and less noisy source location results compared with the time-reversal methods, demonstrating the potential of the method for even global event location.  
\end{itemize}

We first introduce the proposed source imaging framework and key concepts, including PINNs with hard constraints, reference frequency loss function, causality implementation, and data fitting. To demonstrate the effectiveness of the proposed method, Sections~\ref{sec:numerical} and \ref{sec:field} present the numerical examples and field tests. Finally, we discuss the potential of the approach and its limitations in Section~\ref{discussion} and conclude in Section~\ref{sec:conclusion}. 

\section{Theory}
\subsection{Source imaging in the form of wavefield modeling}
The source location imaging problem can be formulated using frequency-domain wavefield modeling, which offers a reduction of dimensionality compared to time-domain wavefield modeling, but more importantly, it will allow us to implement a causal loss function in the depth direction. 
The wave equation in the frequency domain is described by the Helmholtz equation and for a 2D acoustic isotropic medium, it is, 
\begin{equation}
    \frac{\omega^2}{v^2(x,z)}u(x,z,\omega) + \nabla^2{u(x,z,\omega)} = 0,
    \label{gve}
\end{equation}
along with the data boundary condition given by
\begin{equation}
    u(x,z=0,\omega) = D(x, \omega),
    \label{bdc}
\end{equation}
where $u(x,z,\omega)$ is the wavefield corresponding to the frequency $\omega$ and location $(x,z)$, $v(x,z)$ is the velocity, $\nabla^2$ is the Laplacian operator, and $D(x,\omega)$ is the frequency-domain data obtained by applying Fast Fourier transform (FFT) to the recorded data on the surface. 
We use Equations~\ref{gve} and \ref{bdc} to solve for the wavefield and then transform it to the time domain in which the source image corresponds to a slice of the time-domain wavefield where the energy focus is the highest. 
With respect to the source imaging conditions, and for the purpose of this paper, we visually evaluate snapshots of the wavefield, and pick the image with the most focussed source \cite{Artman2010}. 

However, frequency domain modeling for irregular geometry and high frequencies, where fine grids are needed, is memory and computationally intensive. 
It also requires solving for multi-frequencies. 
On the other hand, function learning, by utilizing an NN, is a potential solution in view of its flexibility to an irregular mesh and its instant inference speed. Thus, we use PINNs to represent the frequency-domain wavefield. 

\subsection{Physics-informed neural networks with hard constraints}
In seismic wavefield modeling, physics-informed neural networks (PINNs) can be used to find a neural network function $\Phi(x,z,\omega,\boldsymbol{\theta})$ with parameters $\boldsymbol{\theta}$ that maps the input spatial coordinates and frequency to the value of the complex wavefield that satisfies the Helmholtz equation. 
Using a vanilla PINN often includes two loss terms for each of Equations~\ref{gve} and ~\ref{bdc} to optimize one neural network.
Balancing the contributions of these two terms using a weight affects the convergence and accuracy of the solution. 
The hard constraints \cite{Berg2018,Schiassi2021} concept can alleviate this issue by modifying the representation of the wavefield $u(x,z,\omega)$ to a new form where the boundary condition is inherently satisfied. Specifically, in this paper, we propose to use the following form:
\begin{equation}
u(x,z,\omega) = D(x,\omega) + z\Phi(x,z,\omega,\boldsymbol{\theta}).
\label{hardw}    
\end{equation}
Note that the zero-depth ($z=0$) frequency-domain wavefield, in this case, is always equal to the recorded surface data $D(x,\omega)$ with regardless of the weights of the neural network function $\Phi$. 
By means of the hard constraint, the governing equations are reduced to a single equation given by inserting Equation~\ref{hardw} into Equation~\ref{gve} and~\ref{bdc}, and thus, the loss function of PINNs is reduced to the mean square errors of the resulting single equation.
Thus, the loss function with hard constraints can be written as follows:
\begin{equation}
\begin{aligned}
    \mathcal{L} = \frac{1}{N} \sum_{i=1}^{N}\left|z^i\frac{(\omega^i)^{2}}{(v^i)^2} \Phi(x^{i},z^i,\omega^i,\boldsymbol{\theta})+z^i\nabla^2\Phi(x^i,z^i,\omega^i,\boldsymbol{\theta})+\right.\\
    \left.2\nabla_z\Phi(x^i,z^i,\omega^i,\boldsymbol{\theta})+\frac{(\omega^i)^{2}}{(v^i)^2}D(x^i,\omega^i)+\nabla_x^2D(x^i,\omega^i)\right|_{2}^{2},
    \label{loss}
\end{aligned}
\end{equation}
where $N$ is the number of samples used for training, which includes the samples with variable spatial coordinates $(x^i,z^i)$ and angular frequencies $\omega^i$, $\nabla_x$ and $\nabla_z$ are the gradient operations with respect to $x$ and $z$, respectively.

\subsection{A modified loss function}
Recall that if we do the source imaging in the frequency domain, then the wavefield belonging to multiple frequencies is needed for the transformation to the time domain wavefield. 
As shown in \cite{huang2022single}, direct use of the PINN with the loss (Equation \ref{loss}) would decrease the accuracy and convergence speed. 
Similar to their implementation, we modify the loss function with a single reference frequency loss by replacing $\omega$ with $\alpha \omega_{ref}$ and $u(x,z,\omega)$ with $D(x,\omega) + \alpha z \Phi(x,z,\omega,\boldsymbol{\theta})$, yielding
\begin{equation}
\begin{aligned}
    \mathcal{L} = \frac{1}{N} \sum_{i=1}^{N}\left|\alpha z^i\frac{\omega_{ref}^{2}}{(v^i)^2} \Phi(x^{i},z^i,\omega^i,\boldsymbol{\theta})+\alpha z^i
    \frac{\partial^2\Phi(x^i,z^i,\omega^i,\boldsymbol{\theta})}{\partial(\alpha x)^2}\right.\\
    \left.+\alpha z^i\frac{\partial^2\Phi(x^i,z^i,\omega^i,\boldsymbol{\theta})}{\partial(\alpha z)^2}+2\frac{\partial\Phi(x^i,z^i,\omega^i,\boldsymbol{\theta})}{\partial(\alpha z)}+\right.\\
    \left.
    \frac{\omega_{ref}^{2}}{(v^i)^2}D(x^i,\omega^i)+\frac{\partial^2D(x^i,\omega^i)}{\partial(\alpha x)^2}\right|_{2}^{2},
    \label{sloss}
\end{aligned}
\end{equation}
where $\alpha$ is a scaling factor equal to the ratio of the frequency $\omega$ to the reference frequency $\omega_{ref}$. 
With this scaling factor, the frequency variation is compensated by the scaling of the spatial dimensions, so that the wavelength within the wavefield (to the NN) apparently does not change much with frequency. 
Accordingly, the dimensions of the velocity $v$ are also scaled by $\alpha$ to $v(\alpha x, \alpha z)$. 
For details, we refer the reader to \cite{huang2022single}.

\subsection{Causality loss implementation}
To improve the convergence of PINN, we replace the conventional loss function with a causality loss implementation. 
Since the information guiding the wavefield is coming from the data on the surface, we impose causality to the loss function along the depth axis, so that the wavefield is reconstructed from the earth surface down. 
It is natural to fit the PDE from the shallow to the deep.
This has an equivalence in imaging referred to as downward continuation. 
Like the loss calculation based on the temporal causality \cite{wang2022respecting}, where the causality is with respect to time, here we can apply it with respect to depth. 
In this paper, we are formulating the solution with respect to an initial condition $D(x,\omega)$ (for a solution in depth instead of time). The corresponding modified loss function is given by,
\begin{equation}
\mathcal{L}_c=\frac{1}{N_z} \sum_{i=1}^{N_z} \exp \left(-\epsilon \sum_{k=1}^{i-1} \mathcal{L}\left(z_k, \boldsymbol{\theta}\right)\right) \mathcal{L}\left(z_i, \boldsymbol{\theta}\right),
\label{equ:causality}
\end{equation}
where $\mathcal{L}_c$ is the PDE loss with causality implementation, $\mathcal{L}(z_i,\theta)$ is the loss corresponding to the parameters $\theta$ (equation~\ref{sloss}) at a depth of $z_i$, $N_z$ is the number of samples along the depth axis, and $\epsilon$ is the causality parameter. 
It could be a constant hyperparameter, but we modify it to be a varying one that varies with the number of epochs. 
We use $\epsilon=\epsilon_0 + \frac{\lambda}{epoch + 1}$ as an alternative, where $\epsilon_0$ is the initial value of $\epsilon$ and $\lambda$ is a factor to adjust the change in $\epsilon$.
The causality implies that, for example, the solution at depth 0.5 m depends on the solution at depth 0.3 m, but not the opposite, since the initial condition is at depth zero. 
With this modification, the convergence is faster, and the NN admits better predictions, which we will see later.

\subsection{Fitting the observed data with NNs}
\begin{figure}[!htb]
    \centering
    \includegraphics[width=1.0\columnwidth]{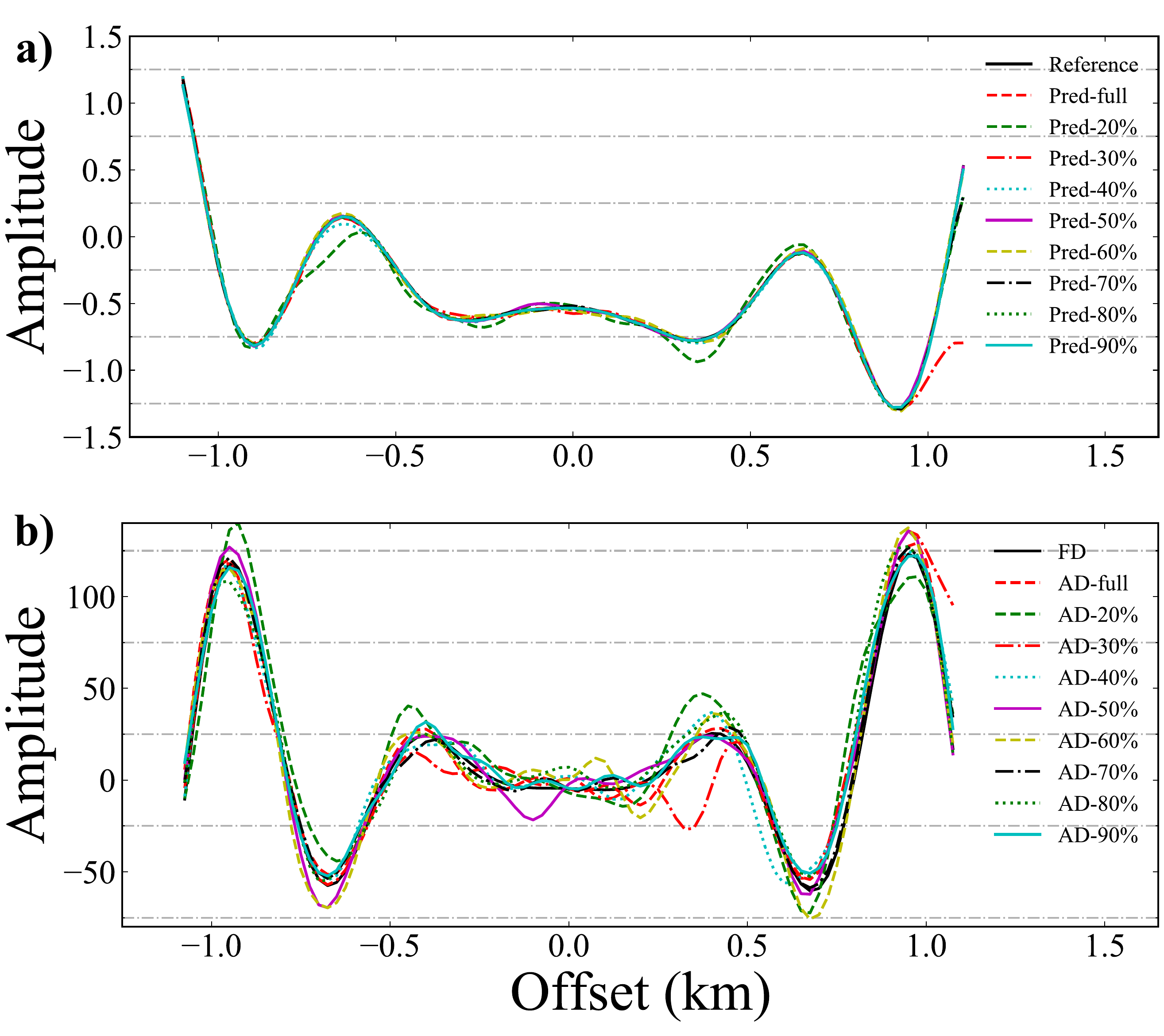}
    \caption{The comparison between the NN function and the 3Hz observed data. a) shows the prediction of NN function trained with 20\% to 90\% of the dataset, and b) is the comparison between the corresponding second-order derivates calculated by AD or the finite-difference method (FD) for various coverage percentages. }
    \label{fig:data_fitting_demo}
\end{figure}

As suggested in \cite{Schiassi2021}, the hard constraint boundary condition could be stored in a neural network to simplify the calculation of the partial derivatives for the hard constraint PINN loss later. 
As a result, we fit a small multi-layer perception (MLP) to the surface observed data in a supervised fashion, so it becomes as an NN function of the horizontal position of the receiver and frequency $D_{\boldsymbol{\theta}}\left(x, \omega\right)$. 
Such training allows for fitting the NN to irregular acquisition (receivers) geometry, which is common in field applications (e.g., the lack of receivers or the non-uniform receiver intervals). 
The size of this data NN function will control the smoothness involved in the interpolation \cite{llanas2006constructive}.
More importantly, it allows for robust derivatives calculation for equation~\ref{sloss} using automatic differentiation (AD). 
Here, the data fitting branch is trained in a supervised manner, where the input of the NN are $x$ and $\omega$, and the output are the real and imaginary parts of the frequency-domain data at that location and frequency. 
Then, the supervised loss function for the data fitting is defined as follows:
\begin{equation}
\mathcal{L}_d(\theta)=\frac{1}{N_r \times N_f} \sum_{j=1}^{N_r \times N_f}\left|D\left(x_j, \omega_j\right)-D_{\boldsymbol{\theta}}\left(x_j, \omega_j\right)\right|_2^2,
\label{data_loss}
\end{equation}
where $N_r$ and $N_f$ are the numbers of samples along the receivers and frequencies. 
Using all samples is not always needed because the noise in the samples may decrease the accuracy, and we usually only need the use of a random subset of the data points to train the NN. 
Here, we show a simple demonstration to highlight the versatility of the approach for data fitting in handling irregularly sampled data. 
We generate data for a small layered model (same as the one used in \cite{Huang2021}) and place 100 receivers on the surface. 
We randomly drop receivers to imitate irregular data to train the network. 
As shown in Figure~\ref{fig:data_fitting_demo}, the prediction looks reasonable even when using only 20\% of the receivers for training. 
The second-order derivative w.r.t $x$ of the NN data function calculated by AD, which is needed for equation~\ref{sloss}, also has good accuracy compared with the second-order derivative calculated using a 5-point central difference formulation or AD with the full data. 

\subsection{The pipeline and the backbone NN architecture}
\begin{figure*}[!htb]
    \centering
    \includegraphics[width=0.75\textwidth]{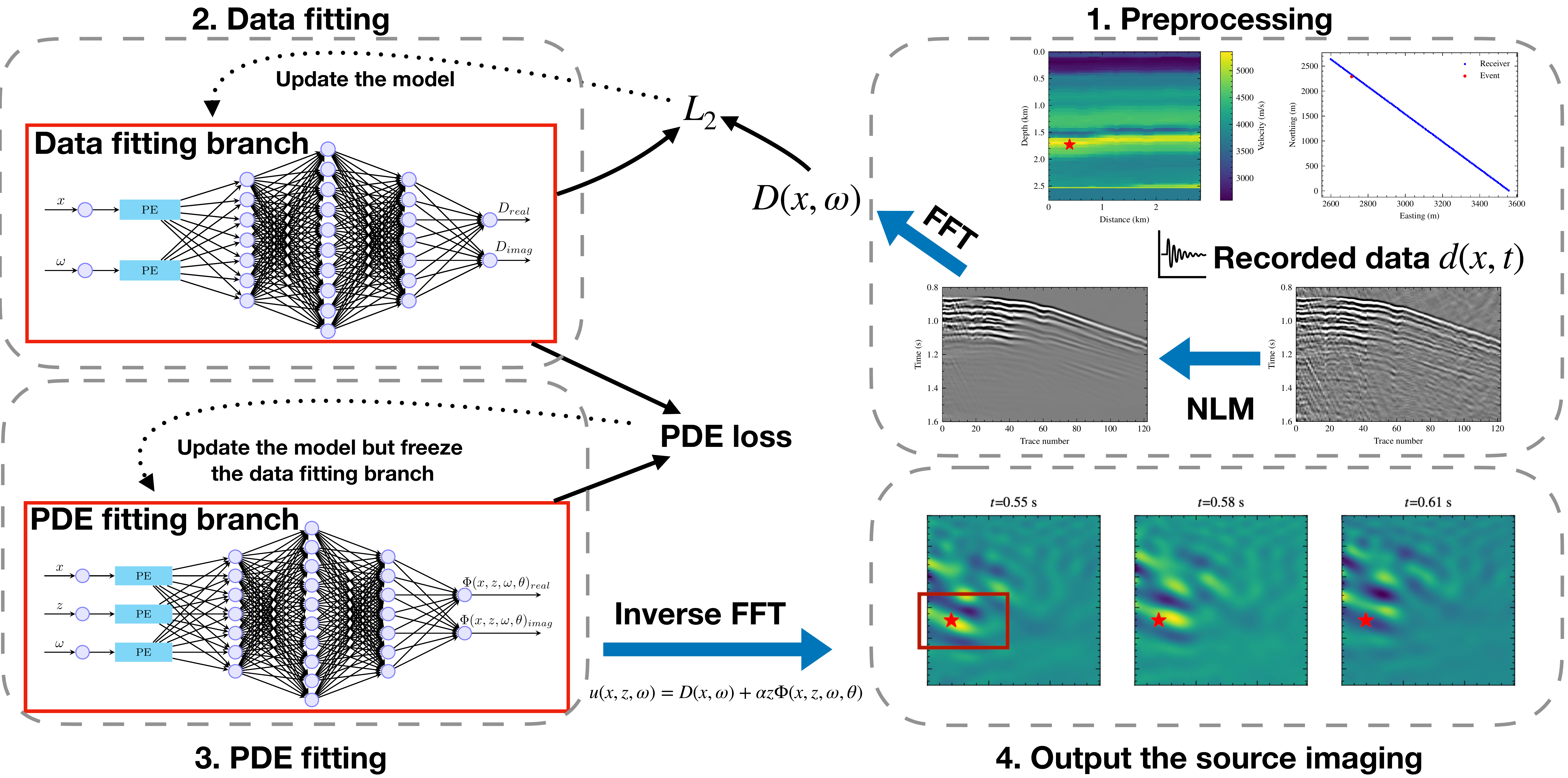}
    \caption{The pipeline of the proposed method. 
    There are two branches: the data fitting branch (\textbf{top}) and the PDE fitting branch (\textbf{bottom}). 
    The recorded data $d(x,t)$ are first transformed to the frequency domain. Then, they are used to train the data NN. Further, the data NN is combined into the PDE fitting branch to train the PDE NN. 
    The output of the PDE NN is transformed to the time-domain snapshots using inverse FFT, where the red box denotes the source imaging where the energy focuses.
    The module PE denotes the positional encoding with sinusoidal functions \cite{Huang2021}.}
    \label{fig:ms-diagram}
\end{figure*}

As shown in Figure~\ref{fig:ms-diagram}, the preprocessed (e.g., non-local means filtering, NLM) recorded data on the surface are first transformed into the frequency domain via FFT and then used to train a small NN for data fitting. 
This trained NN will be used for the evaluation of the data term and calculation of the second-order derivate by AD in the partial differential equation (PDE) fitting. 
In the PDE fitting branch, the parameters of the data fitting branch are frozen, and another new larger NN is used to fit the PDE. 
To accelerate the convergence, we use the causality implementation of the PDE loss. 
We show more details of the training pipeline in Algorithm~\ref{alg}.
After training, we predict the multi-frequency wavefield using the NN and utilize inverse FFT to get the time-domain snapshot. 
The locations where the energy focus are the source locations.
\begin{algorithm}[!htb]
Collected $N_r\times N_f$ recorded data on the surface $\{D(x_j,\omega_j)\}^{N_r\times N_f}_{j=1}$, and $\{(x_i,z_i,\omega_i)\}^{N}_{i=1}$ \\
Initiate: NN parameters $\boldsymbol{\theta}_1$ and $\boldsymbol{\theta}_2$\\
\For{each epoch in the data fitting}{
    Input: $(x_j,\omega_j)$ and corresponding labels $D(x_j,\omega_j)$\\
    Output: $D_{\boldsymbol{\theta}_1}(x_j,\omega_j)$\\
    Calculate the loss function of equation~\ref{data_loss}\\
    Update: NN parameters $\boldsymbol{\theta}_1$
}
Freeze the NN parameters $\boldsymbol{\theta}_1$\\
\For{each epoch in the PDE fitting}{
    Input: $(x_i,z_i,\omega_i)$\\
    Output: $\Phi(x_i,z_i,\omega_i,\boldsymbol{\theta}_2)$\\
    Evaluate $D_{\boldsymbol{\theta}_1}(x_i,\omega_i)$ and replace it in equation~\ref{sloss}\\
    Calculate the loss function of equation~\ref{sloss}\\
    Update: NN parameters $\boldsymbol{\theta}_2$
}
\caption{The training pipeline}
\label{alg}
\end{algorithm}

As for the implementation of the neural networks used in this pipeline, we utilize positional encoding \cite{Huang2021} to help the network deal with the complex wavefield variations and help the convergence of the NN. The backbone of the NN is Multilayer Perception (MLP \cite{longstaff1987pattern}) with sine as the activation function. The size of the networks and training details will be shared in the examples.  

\section{Numerical examples}
\label{sec:numerical}
We first evaluate the proposed method on synthetic data, including single-source event and multiple-source events cases.
The synthetic data are generated for the 2D slice from the 3D Overthrust model \cite{Aminzadeh1996ThreeDS} (Figure~\ref{fig:single-source-v}a). The model size is 501$\times$161 with a spatial interval of 25 m. 
During the training for synthetic examples, we use an MLP with two hidden layers, with 128 and 16 neurons in the layers from shallow to deep, respectively, to fit the data, but we reduced the size of MLP to one hidden layer of 128 neurons for samples less than 40. 
While for the PDE fitting, we use a larger MLP with six hidden layers, with 256, 256, 128, 128, 64, and 64 neurons from shallow to deep, respectively. We also include the positional encoding for $x$, $z$, and $\omega$ with a level of 4. The frequency range for both tasks is from 3 to 12 Hz and we choose 12 Hz as the reference frequency.

\subsection{A single-source event case}
\begin{figure}[!htb]
    \centering
    \includegraphics[width=0.186\textwidth]{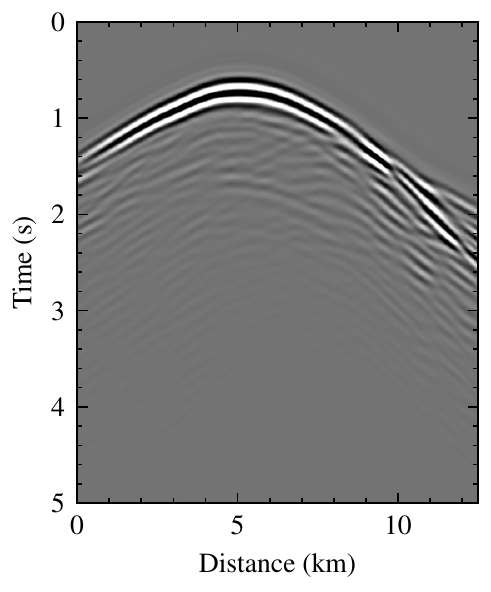}
    \caption{The filtered (3-12 Hz) recording of one source event (located at (5.0, 3.0) km) with receivers covering the whole surface.}
    \label{fig:record_1s}
\end{figure}

\begin{figure}[!htb]
    \centering
    \includegraphics[width=0.45\textwidth]{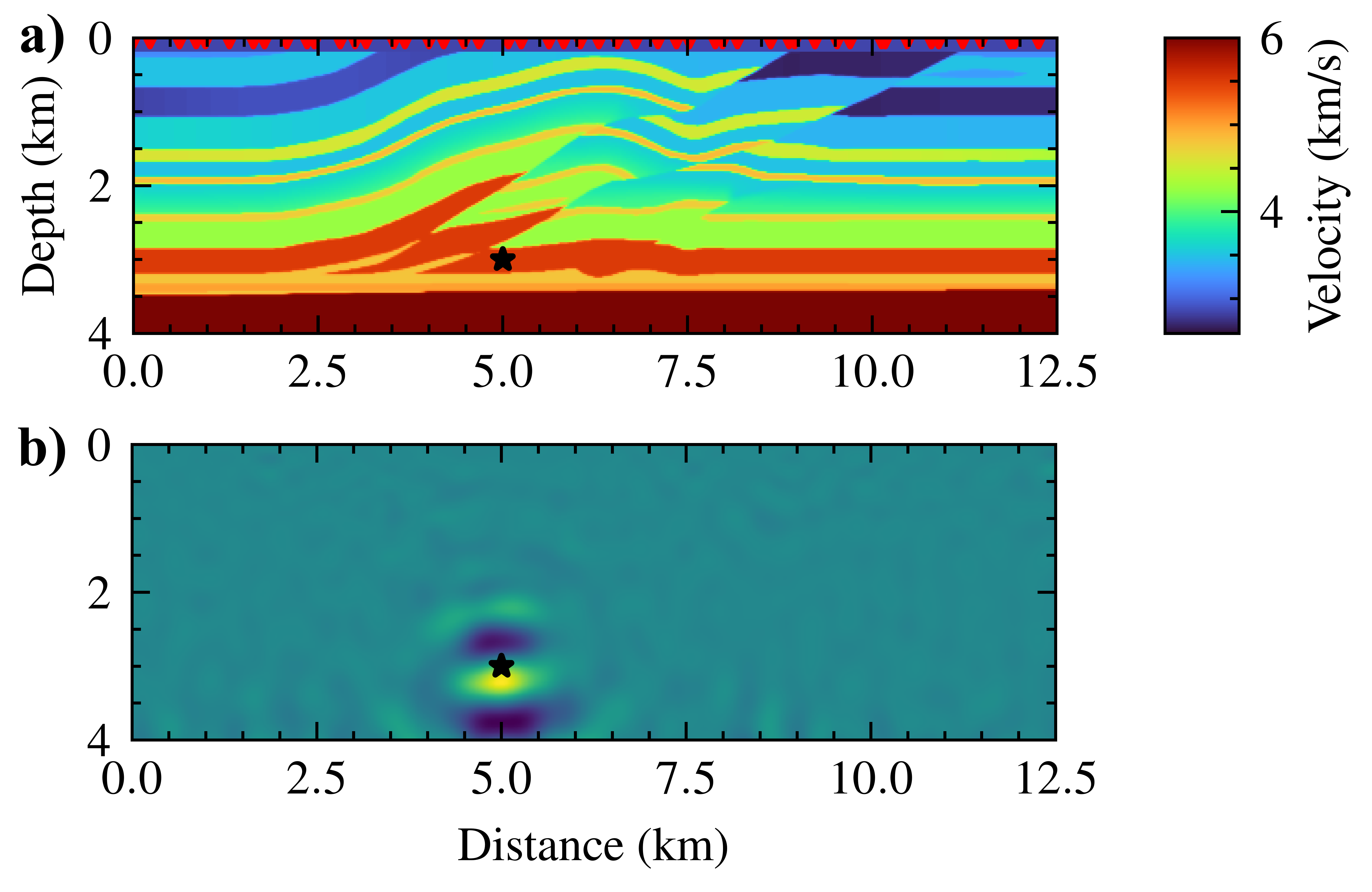}
    \caption{a) shows the Overthrust velocity model, the source event location (denoted by the black star), and the random receivers (denoted by the red triangles);
    b) is the source image courtesy of the proposed method.}
    \label{fig:single-source-v}
\end{figure}

\begin{figure}[!htb]
    \centering
    \includegraphics[width=0.37\textwidth]{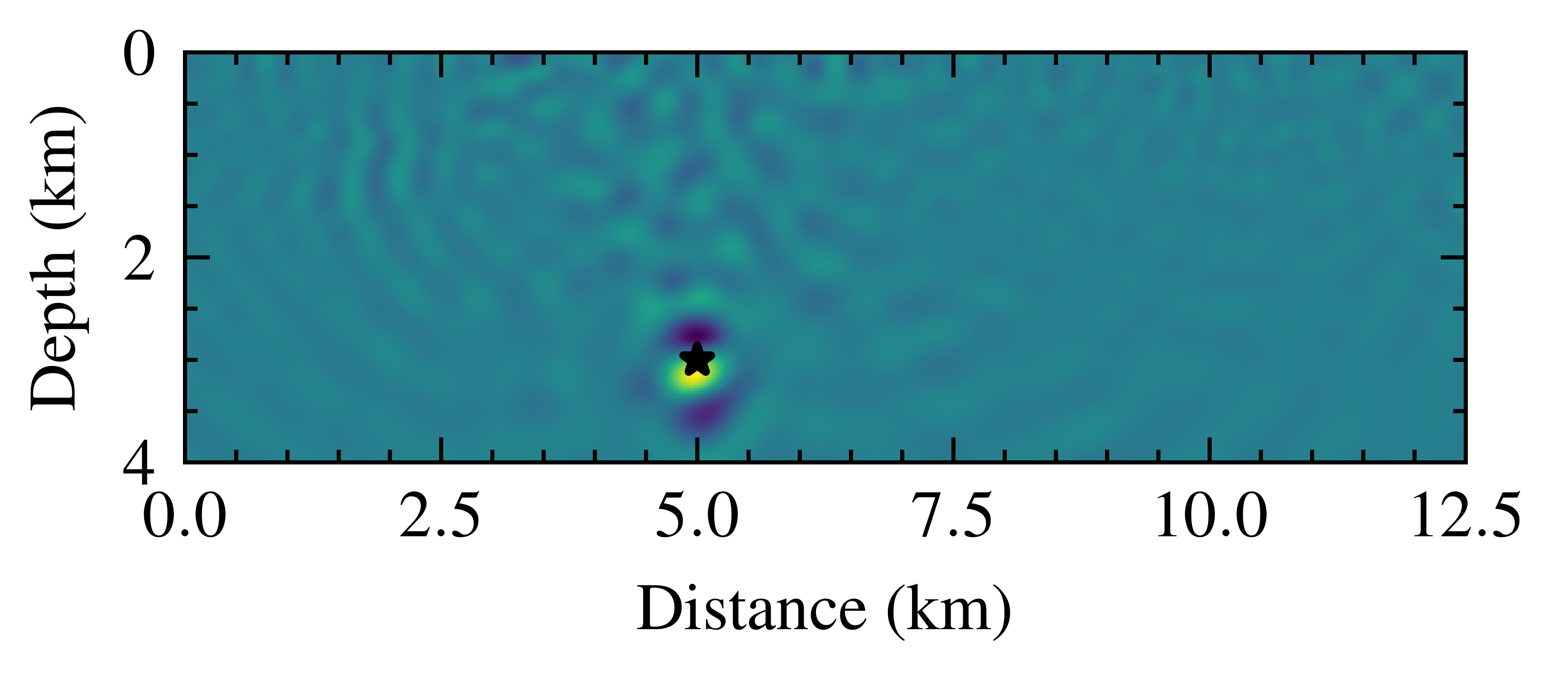}
    \caption{The source image courtesy of a numerical finite-difference solver.}
    \label{fig:1s_numerical}
\end{figure}

In this subsection, we first test the proposed method in a toy problem, that has only one source event. A source is placed at (5.0, 3.0) km. 
We first do the forward modeling on the Overthrust model due to this source and recorded the data (Figure~\ref{fig:record_1s}) with 50 randomly placed receivers on the surface, shown in Figure~\ref{fig:single-source-v}a. 
We train the NN for data fitting with positional encoding and train for 20000 epochs using an Adam optimizer with a learning rate of 1e-3. 
Then we train another NN for PDE fitting with 80000 random samples for 6000 epochs, where we freeze the parameters of the NN for data fitting. 
We evaluate the NN on a uniform grid with 501$\times$161 grid points and an interval of 25 m to obtain the multi-frequency wavefields.
As we have the source information for this case, we directly obtain the source image by the summation of the wavefield over frequencies (equal to the time-domain snapshot at zero time). 
As shown in Figure~\ref{fig:single-source-v} b, the source location in the image is consistent with the ground truth (black star in Figure~\ref{fig:single-source-v}a), which demonstrates the accuracy of the proposed method. 
We also applied the time reversal imaging using the finite-difference method to obtain the source image as a baseline for comparison, as shown in Figure~\ref{fig:1s_numerical}. 
Since we ignited the source at time zero, we evaluate the image at time zero, which is equivalent to the direct summation of wavefields over all frequencies.   
It is obvious that with 50 receivers, both methods could reconstruct the source image, while the result of the proposed method is cleaner compared to the numerical method, which contains the artifacts due to relative-sparse sampling.
Next, we test more complex scenarios.

\subsection{A more complex case including triple-sources event}
In a more realistic scenario, there might be multiple sources distributed at variable locations in the subsurface. 
Thus, we test the proposed method in this more complex situation.
\begin{figure}[!htb]
    \centering
    \includegraphics[width=0.186\textwidth]{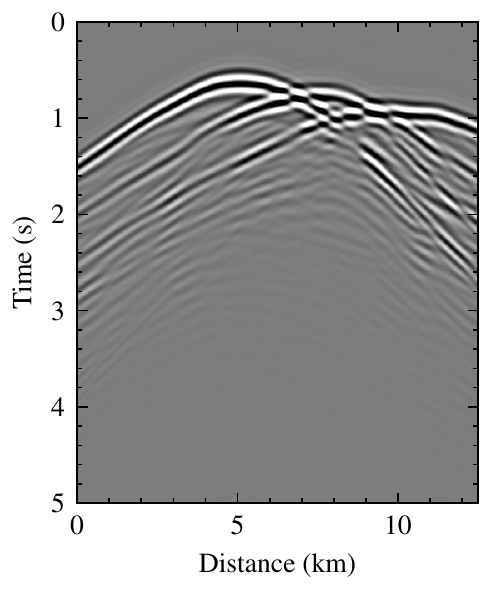}
    \caption{The filtered (3-12 Hz) recording of triple-sources event (located at (5.0, 2.5), (7.5, 3.0), and (10.0, 3.5) km) with receivers covering the whole surface.}
    \label{fig:record_3s}
\end{figure}
\begin{figure*}[!htb]
    \centering
    \includegraphics[width=0.74\textwidth]{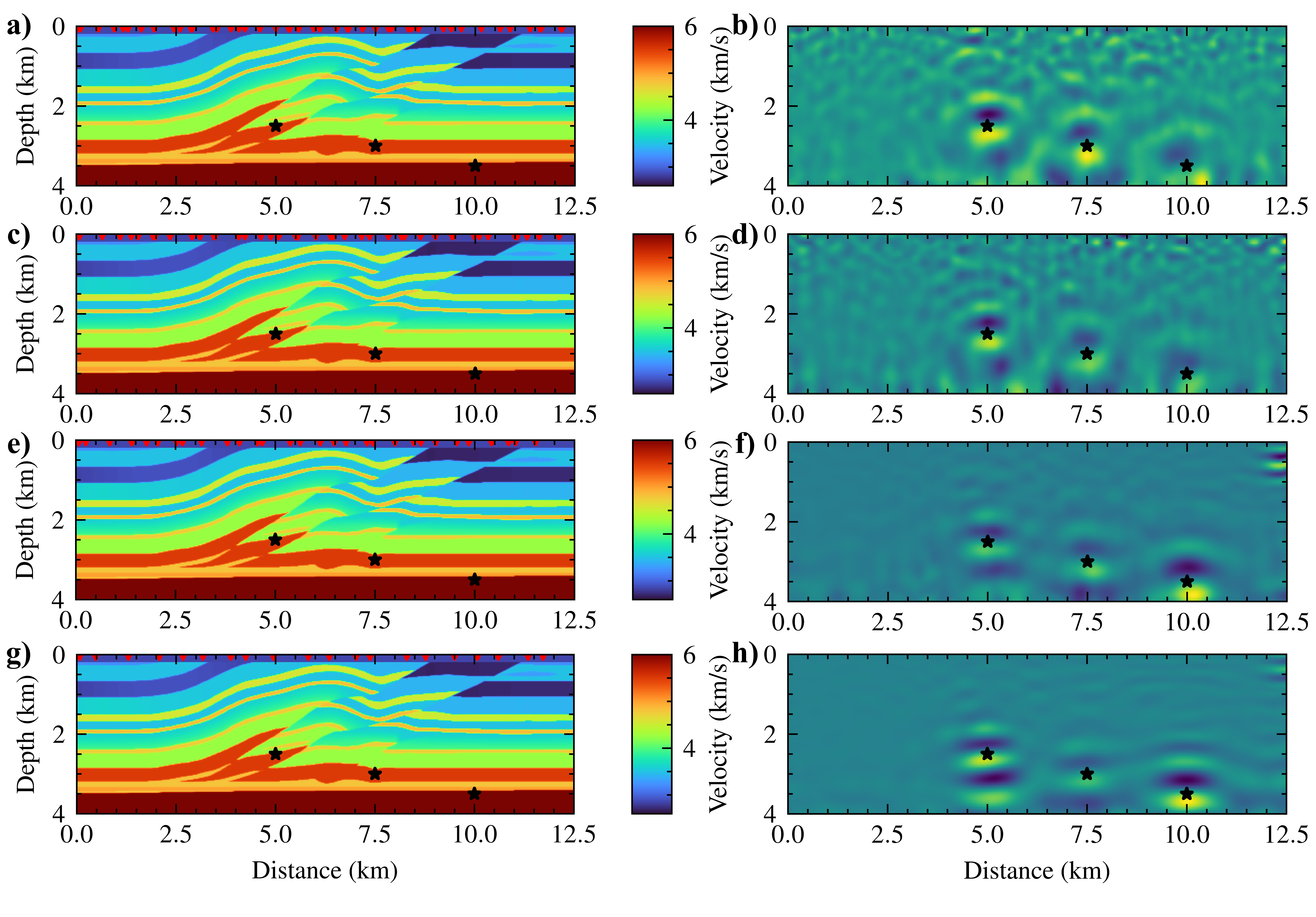}
    \caption{The source imaging results using various numbers of random receivers, specifically 50 (a), 40 (c), 30 (e), and 20 (g) receivers denoted by the red triangles on top of the velocity models (near the surface). 
    The sources denoted by the black stars are distributed at different locations: $(5.0,2.5)$, $(7.5,3.0)$, and $(10.0,3.5)$ km. 
    The source imaging results corresponding to various numbers of receivers by the proposed method are shown in b), d), f), and h).}
    \label{fig:triple-source}
\end{figure*}
\begin{figure}[!htb]
    \centering
    \includegraphics[width=0.37\textwidth]{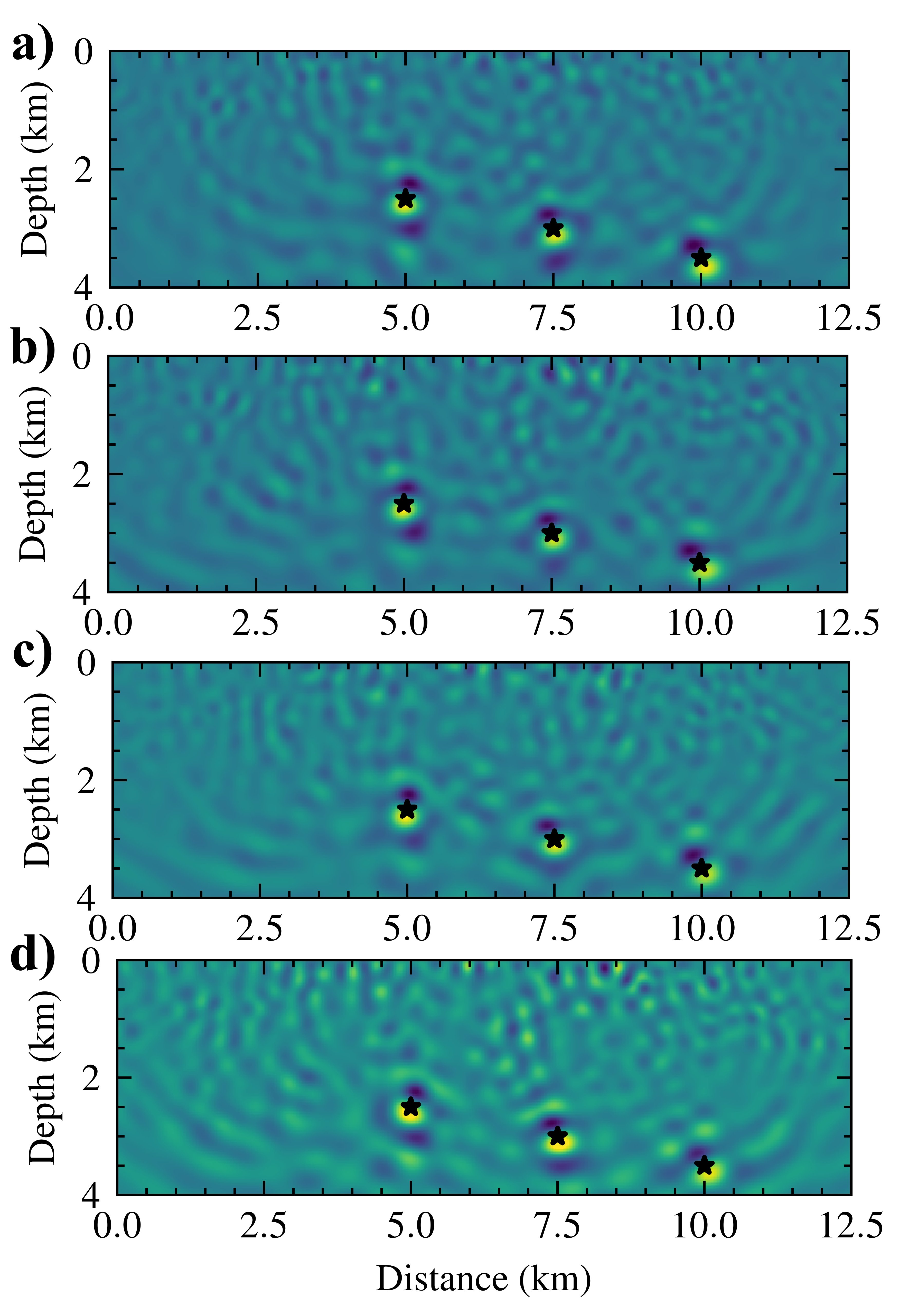}
    \caption{The source imaging results by the numerical time reversal imaging with the finite-difference method using various numbers of random receivers, specifically 50 (a), 40 (b), 30 (c), and 20 (d) receivers denoted by the red triangles in Figure~\ref{fig:triple-source}.}
    \label{fig:triple-source-numerical}
\end{figure}
The black stars in Figure~\ref{fig:triple-source}a show the source locations, and we use forward modeling from these sources to obtain the data (Figure~\ref{fig:record_3s}), recorded by a varying number of random receivers. 
With the same hyperparameters and training process mentioned in the last section, we tested the proposed method with data received at 50 (Figure~\ref{fig:triple-source} a), 40 (Figure~\ref{fig:triple-source} c), 30 (Figure~\ref{fig:triple-source} e), and 20 (Figure~\ref{fig:triple-source} g) random receivers. 
As shown in Figures~\ref{fig:triple-source}b,d,f, and h, the source locations are accurately extracted by the proposed method.
Even in the case of only 20 random receivers placed along the 12.5 km distance, we can still get focused source imaging results.
While for the numerical time reversal imaging using the finite-difference method, shown in Figure~\ref{fig:triple-source-numerical}, the result is not as clean as those by the proposed method, especially for the area near the source.
Comparing our proposed method to numerical time reversal imaging, although our results seem slightly smoothed, they remain sufficient for identifying source locations. In fact, the smoothness can assist in the automatic detection of the source location.
This demonstrates the capability of the proposed method in handling sparse and irregular observations, showing good potential for the application to field data. 

\begin{figure*}[!htb]
    \centering
    \includegraphics[width=0.76\textwidth]{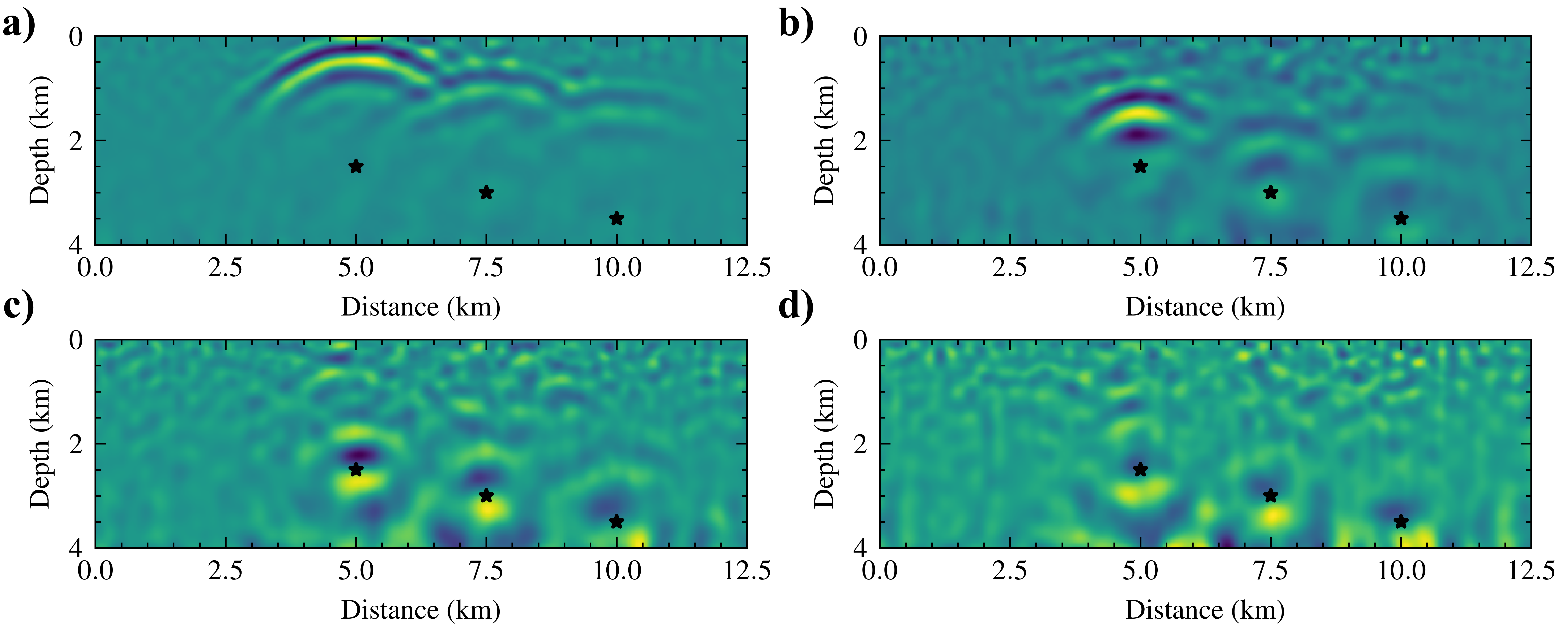}
    \caption{Time-domain snapshots of wave propagation at various times, e.g., a) 5.50 s, b) 5.25 s, c) 5.00 s, and d) 4.75 s, where the energy of the wavefield focus best at 5.0 s (c).}
    \label{fig:triple-source-time-domain}
\end{figure*}

But before that, we need to test the performance of the proposed method in the passive seismic monitoring scenario, as this is a more common case. 
The above experiments are built on an active seismic scenario in which the source is ignited at zero time.
For passive seismic data, we often do not know the source ignition time. 
Thus, instead of the direct summation of the wavefields over frequencies, we inverse Fourier transform the wavefields to obtain time-domain snapshots.
The snapshot that shows the largest focusing of energy is often regarded as the ignition time of the source \cite{Artman2010}.
Thus, with a similar setting to the previous experiments, we execute the forward modeling from the triple sources but ignite the source at 5 s, which we consider as an unknown.
We use the proposed pipeline to obtain the time-domain snapshots, and we show several snapshots from 4.75 to 5.5 s in Figure~\ref{fig:triple-source-time-domain}. 
We observe that the energy focuses best at the snapshot at 5.00 s, which is consistent with the ignition time. The locations where the energy focus are also consistent with the ground truth of the source locations.

\section{The field test to the hydraulic fracturing data}
\label{sec:field}
\begin{figure}[!htb]
    \centering
    \includegraphics[width=0.98\columnwidth]{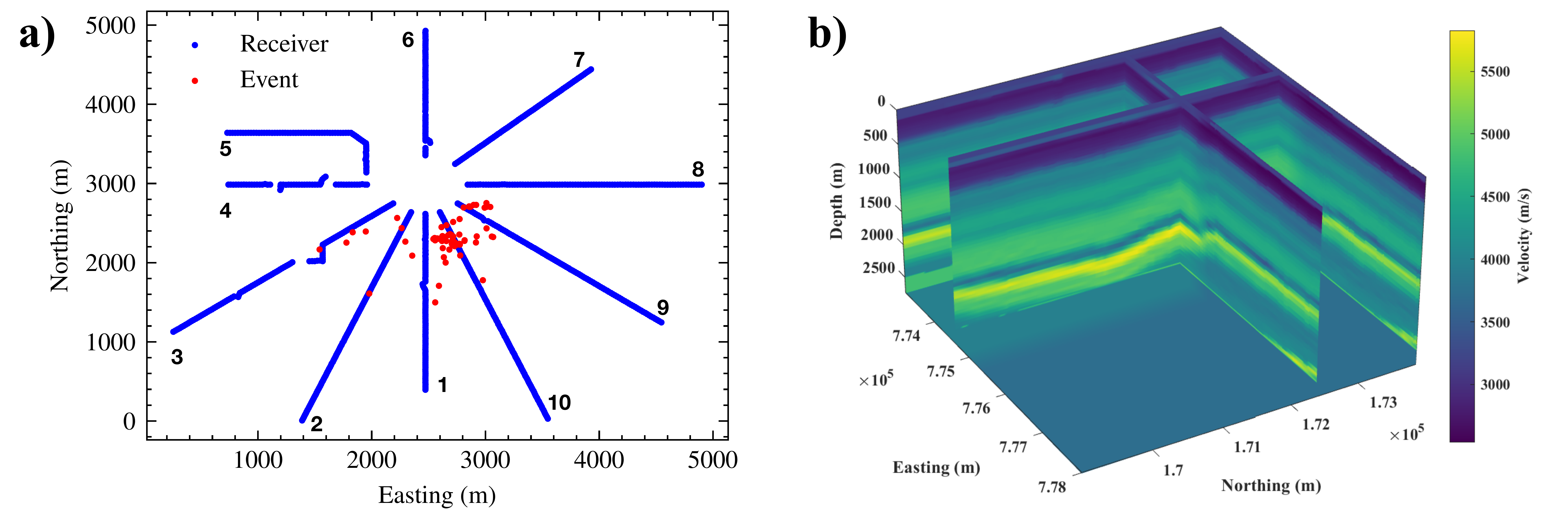}
    \caption{Passive seismic monitoring in Oklahoma, a) is the recording geometry in this region where the blue dots denote the receivers, and red dots denote the locations of the previously estimated events projected on the 2D plane; b) is the corresponding 3-D P-wave velocity.}
    \label{fig:3d_field_vel}
\end{figure}

\begin{figure}[!htb]
    \centering
    \includegraphics[width=0.98\columnwidth]{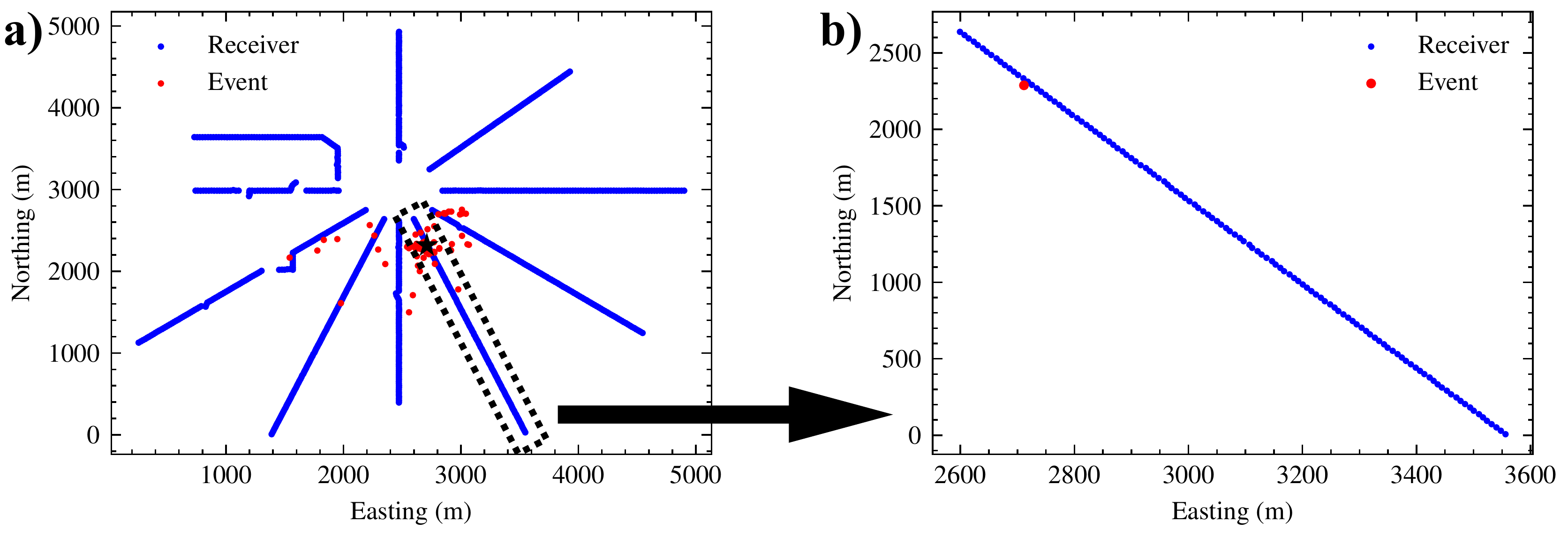}
    \caption{The selected event 22 and the recording line. a) is the original geometry, where the black box denotes the selected recording line, and the black star denotes the location of the event; b) is the zoom of the corresponding geometry.}
    \label{fig:field_line10_event8}
\end{figure}

\begin{figure}[!htb]
    \centering
    \includegraphics[width=0.98\columnwidth]{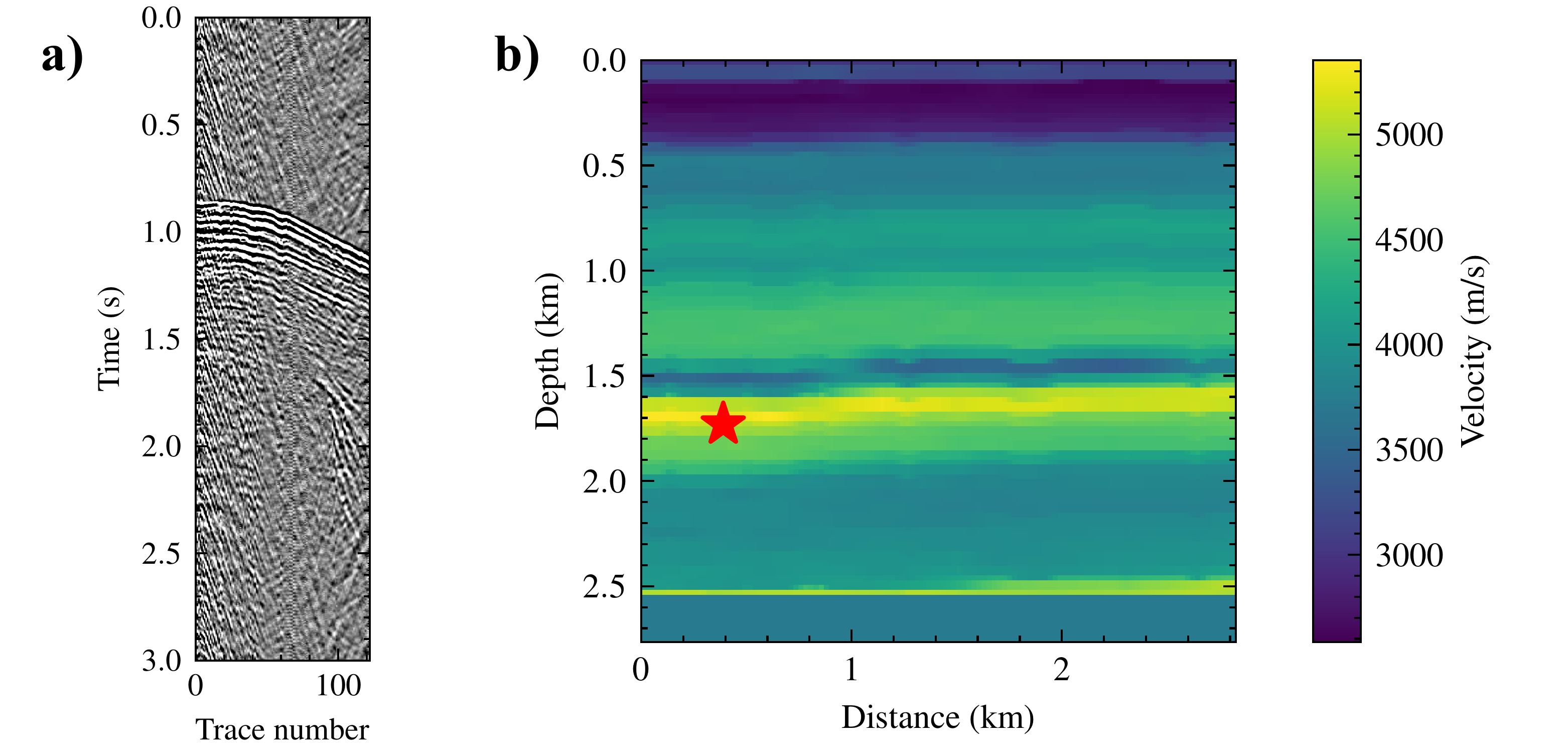}
    \caption{The original recorded data for a single event a) and the 2D P-wave velocity profile b), where the source location provided by the data provider is denoted by a red star.}
    \label{fig:field_line10_event8_vel_record}
\end{figure}

The above synthetic examples have demonstrated the potential for a field data application. 
Thus, in this section, we have further tested on the field vertical-component data acquired in the Arkoma Basin in North America. 
The data are recorded through passive seismic monitoring during a hydraulic fracturing stimulation for a shale gas reservoir \cite{anikiev2014joint,stanvek2017seismicity,Wang2022}. 
Figure~\ref{fig:3d_field_vel}a shows the whole region covering a square area of 5$\times$5 km$^2$ with a maximum depth of 2.75 km. 
There were 75 events (red dots) captured in the four days of monitoring and recorded by 911 receivers (blue dots) on the surface. 
The 3D P-wave velocity (Figure~\ref{fig:3d_field_vel}b) is obtained from an active seismic experiment in the region and adjusted by the well-log information \cite{anikiev2014joint}. 
In this case, we do not have the exact zero-tag information of the recordings. 
Here, to demonstrate the performance of the proposed method on the field data, we select two events based on their signal-to-noise ratio of the recorded event and the geometry of the line and the event, so it complies with the 2D implementation of the proposed method.
The first event is extracted from the recordings along line 10, denoted by the black box, shown in Figure~\ref{fig:field_line10_event8}. The corresponding 2D velocity profile and recorded data for the event are shown in Figure~\ref{fig:field_line10_event8_vel_record}. Prior to applying our PINN imaging, we use several pre-processing steps, including the Non-local means filter \cite{buades2005non} and a bandpass filter on the data. 
\begin{figure}[!htb]
    \centering
    \includegraphics[width=0.98\columnwidth]{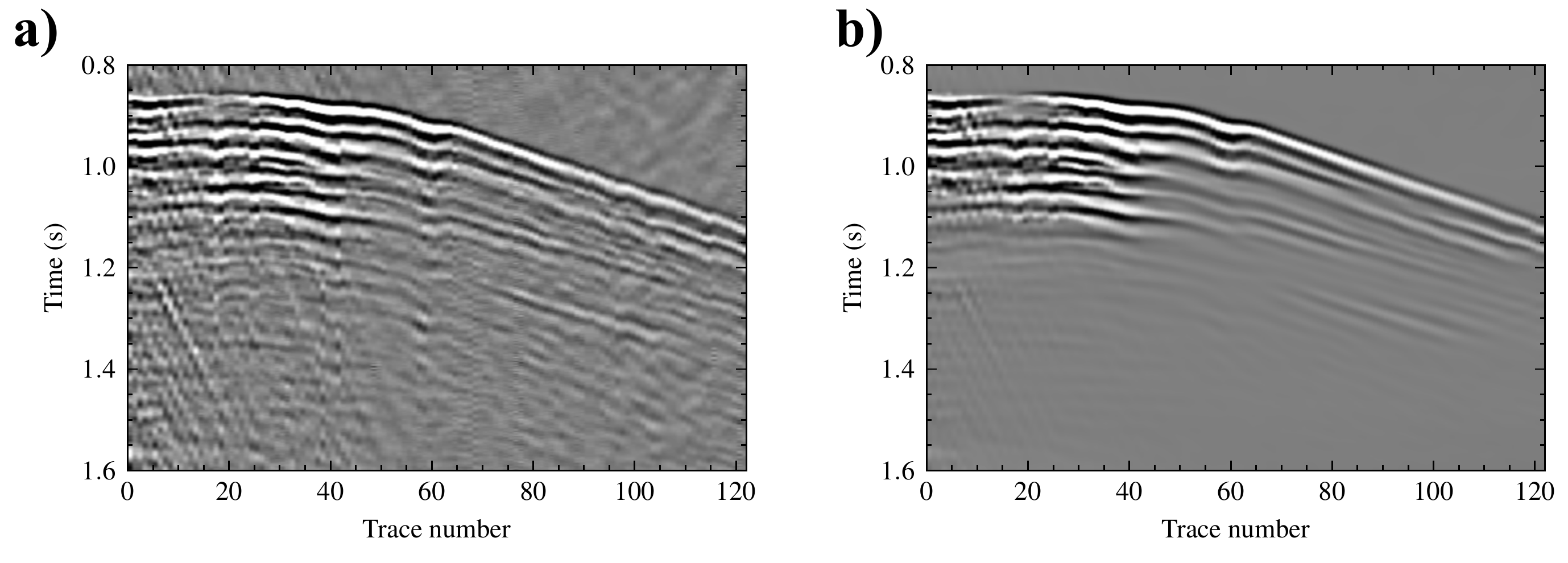}
    \caption{The recording of event 22, a) is the original recording with recording time from 0.8 s to 1.6 s; b) is the corresponding processed data from 3 Hz to 12 Hz after non-local means filtering where the patch size used for denoising is 15, maximal distance in pixels where to search patches used for denoising is 21, and the cut-off distance is 2.0.}
    \label{fig:field_line10_event8_processing}
\end{figure}
\begin{figure}[!htb]
    \centering
    \includegraphics[width=0.98\columnwidth]{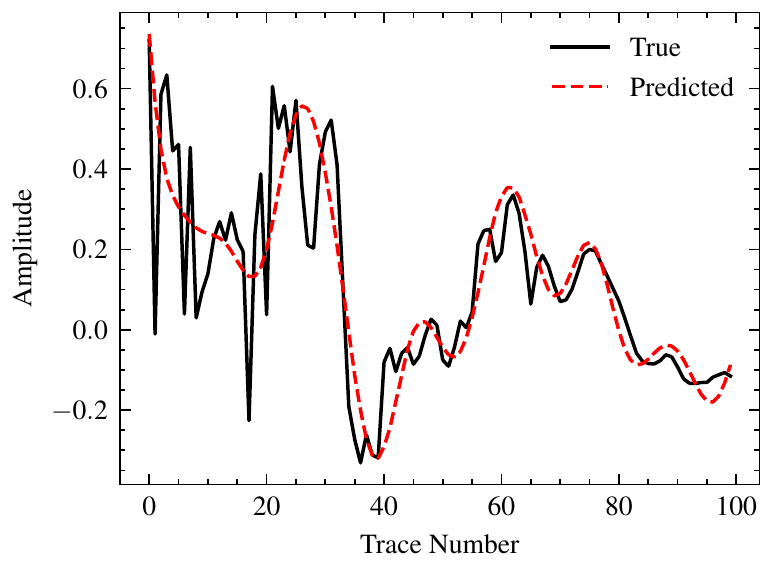}
    \caption{A comparison between the data fitting result and recorded data of 3Hz for the first 100 receivers. The black line denotes the recorded data, while the red line is the predicted data by NN.}
    \label{fig:field_line10_event8_fitting}
\end{figure}

After preprocessing, we start the proposed pipeline (Figure~\ref{fig:ms-diagram}). We transform the recorded data to the frequency domain and train an NN of size \{18, 128, 2\} with positional encoding where L=4 to fit the 3-12Hz recorded data on the surface. As we claimed before, the data fitting network could deal with a big part of the receivers missing. Here, we randomly pick 70 receivers out of 122 for the training. The benefit of this coarse fitting (Figure~\ref{fig:field_line10_event8_processing}) is to allow for a smooth representation of the data that captures the key moveout information. After 4000 epochs of supervised training with an Adam optimizer and a learning rate of 1e-3, the predicted data are shown in Figure~\ref{fig:field_line10_event8_fitting}. We could obviously observe that the main features of the original data are well-reconstructed by the neural network, granted we ignore the sharp changes, which could be due to noise (wavefields are inherently smooth). 
\begin{figure}[!htb]
    \centering
    \includegraphics[width=1.0\columnwidth]{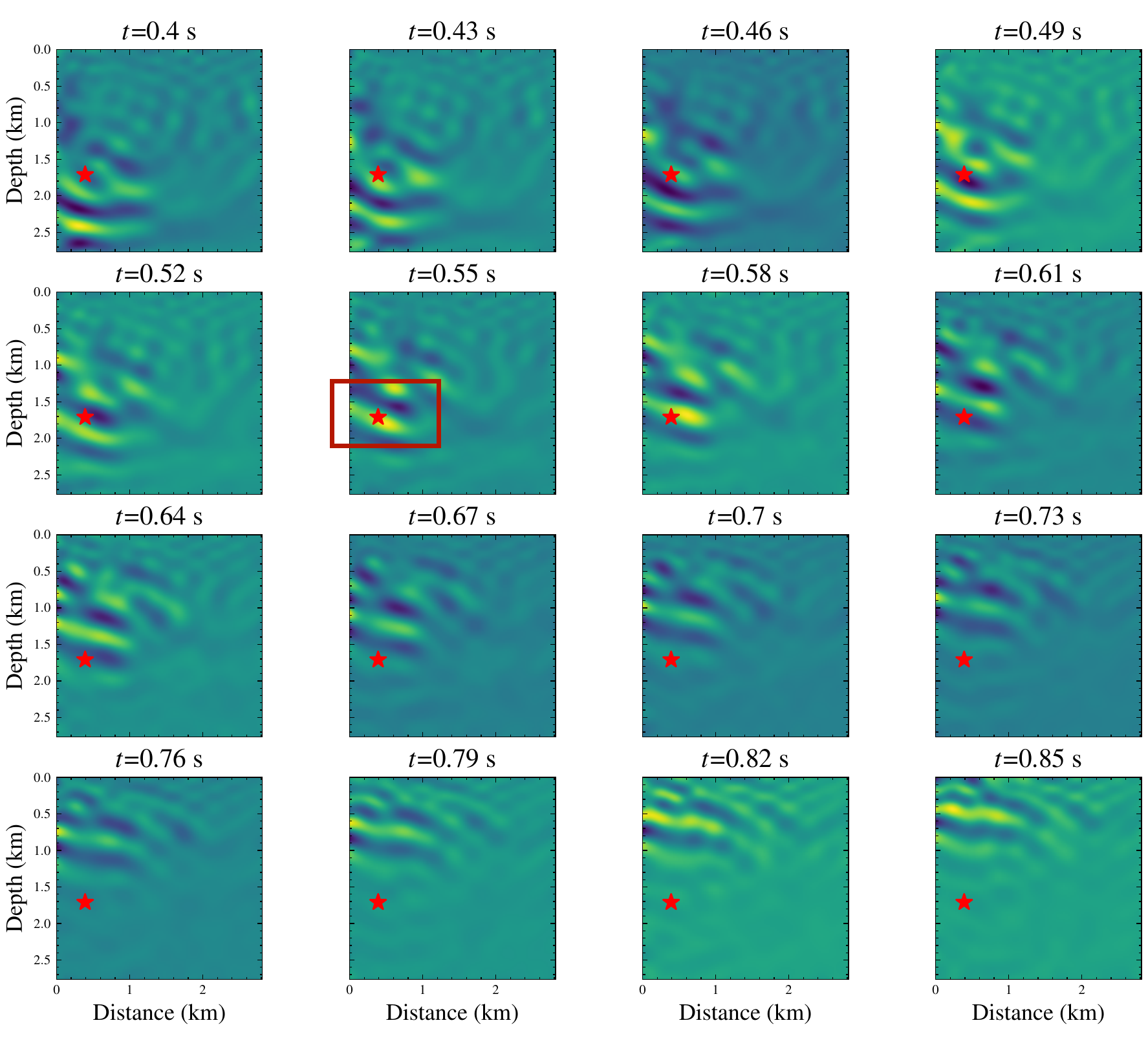}
    \caption{The time-domain snapshots of the wavefield at various times with the proposed method. It seems that the wavefield focus best at 0.55 s. The red box denotes the location where the energy is focused, which is consistent with the provided location (red star).}
    \label{fig:field_line10_event8_td_image}
\end{figure}

\begin{figure}[!htb]
    \centering
    \includegraphics[width=1.0\columnwidth]{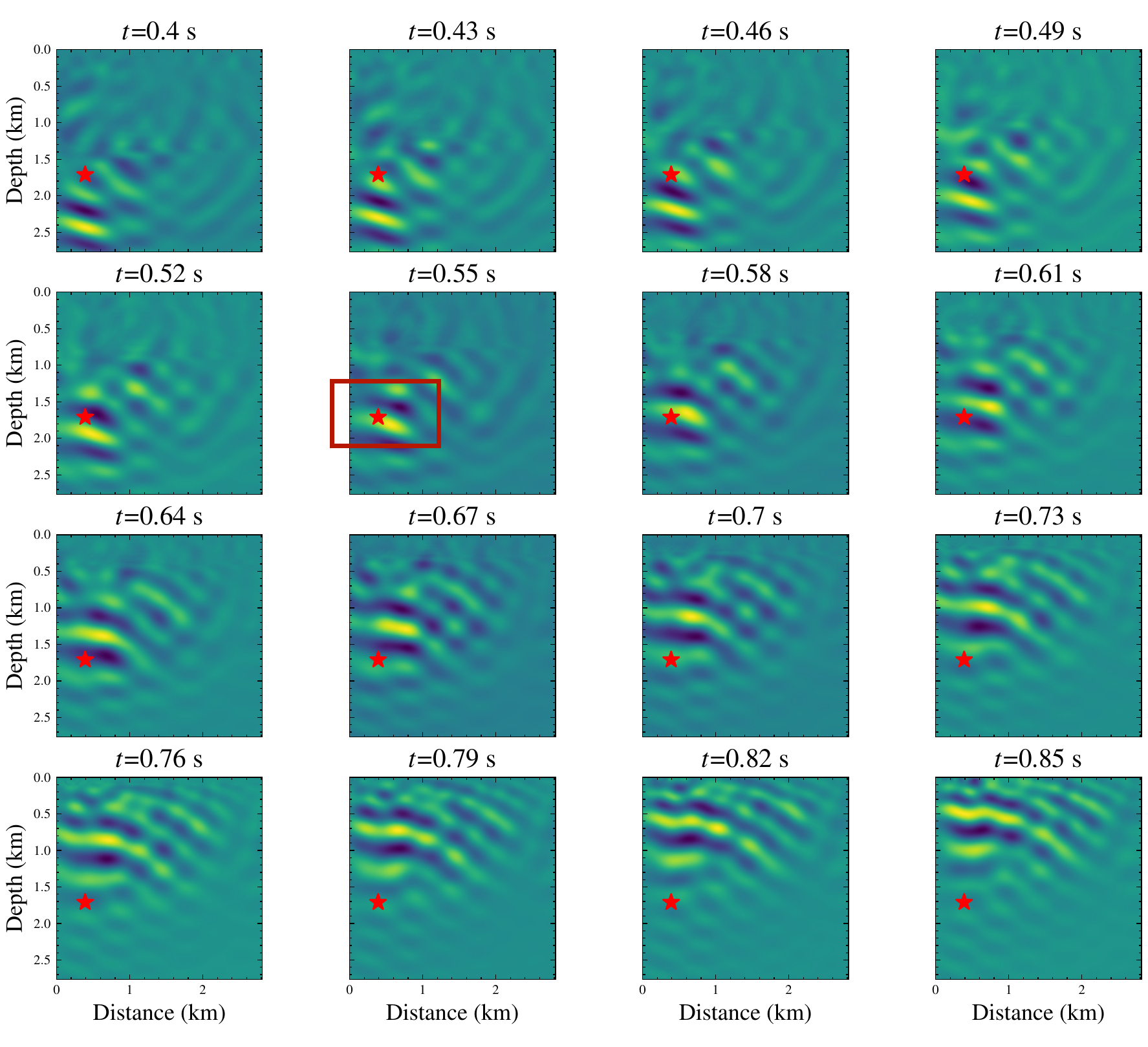}
    \caption{The time-domain snapshots of the wavefield at various times using the numerical time reversal imaging with finite-difference method. The red stars denote the reference source location.}
    \label{fig:field_line10_event8_td_image_numerical}
\end{figure}

Then, the next step is to perform the PDE fitting branch in Figure~\ref{fig:ms-diagram}. Here, we use a larger NN of size \{27, 256, 256, 128, 128, 64, 64, 2\} with positional encoding where L=4 to ensure the capacity of the NN is enough to represent this high-dimensional wavefield. The loss function here is the causality implementation of equation~\ref{sloss}. After training with an Adam optimizer using a learning rate of 1e-3 for 1500 epochs, we obtain the frequency-domain wavefield. An inverse Fourier transform provides time-domain snapshots (source imaging). Figure~\ref{fig:field_line10_event8_td_image} shows these snapshots in which the source image seems to focus on the right source location consistent with the provided location at the estimated time of around 0.55 s. 
Meanwhile, we also performed the numerical time reversal imaging on the filtered field data after preprocessing, and the imaging results are shown in Figure~\ref{fig:field_line10_event8_td_image_numerical}.
We found that the time-reversal result is not as clean or crisp as the result of our proposed method due to the numerical errors from the irregular receiver spacing on the surface as well as the noise and discontinuity in the field data, which can be smoothed out by function learning in the proposed method.

\begin{figure}[!htb]
    \centering
    \includegraphics[width=0.98\columnwidth]{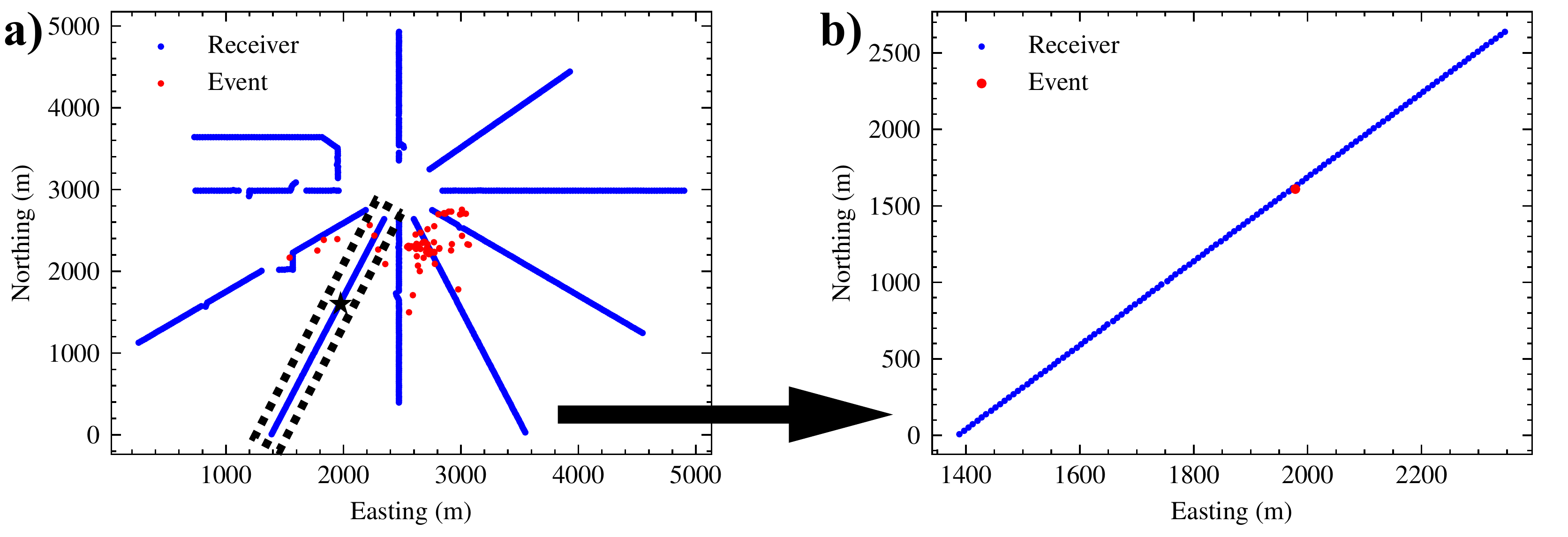}
    \caption{The selected event 62 and the recording line 2. a) is the original geometry, where the black box denotes the selected recording line, and the black star denotes the location of the event; b) is a zoom of the corresponding geometry.}
    \label{fig:field_line2_event62}
\end{figure}

\begin{figure}[!htb]
    \centering
    \includegraphics[width=0.98\columnwidth]{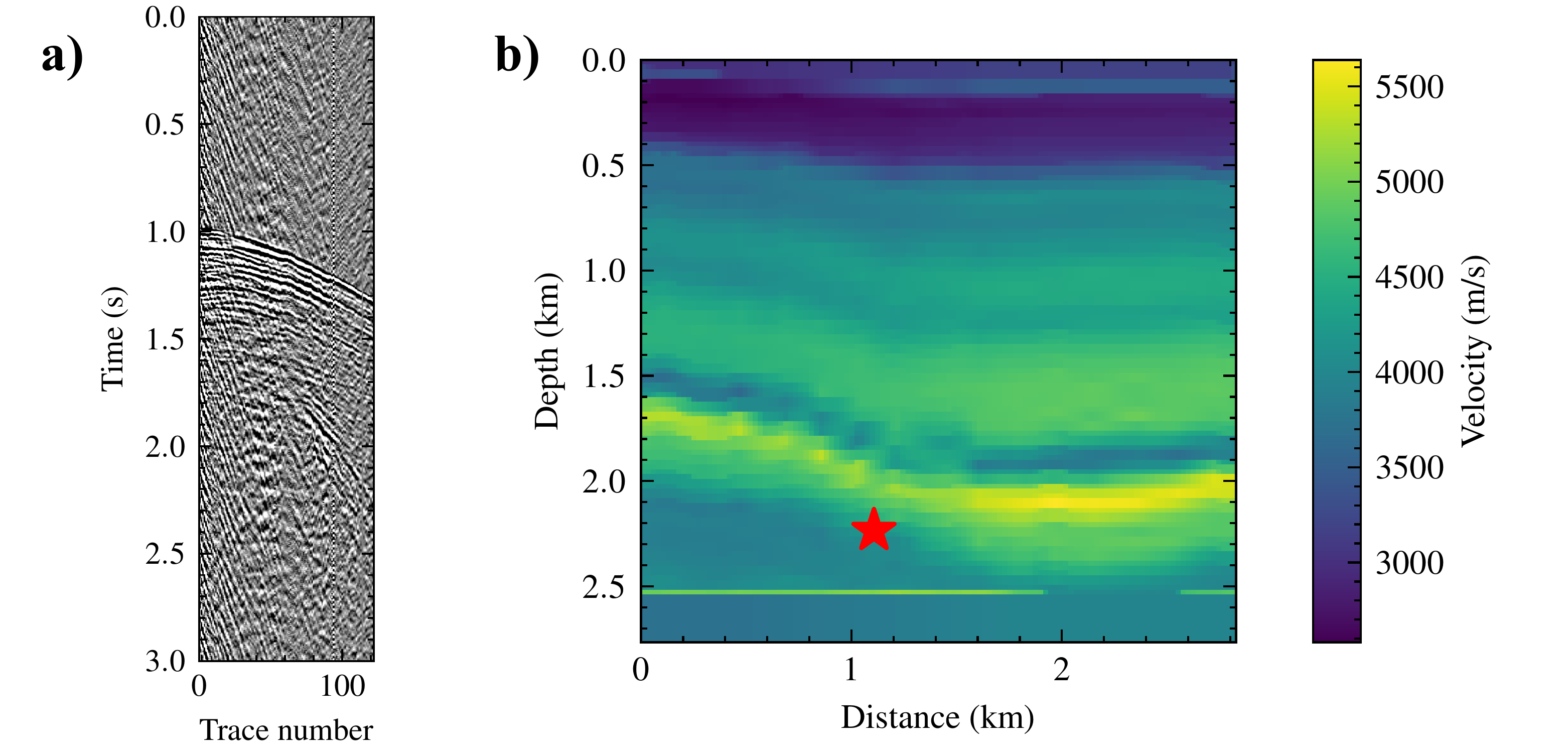}
    \caption{The original shot gathers of line 2 due to event 62, a) and the 2D P-wave
    velocity profile b), where the red star denotes the source location given by the data provider.}
    \label{fig:field_line2_event62_vel_record}
\end{figure}

\begin{figure}[!htb]
    \centering
    \includegraphics[width=1.0\columnwidth]{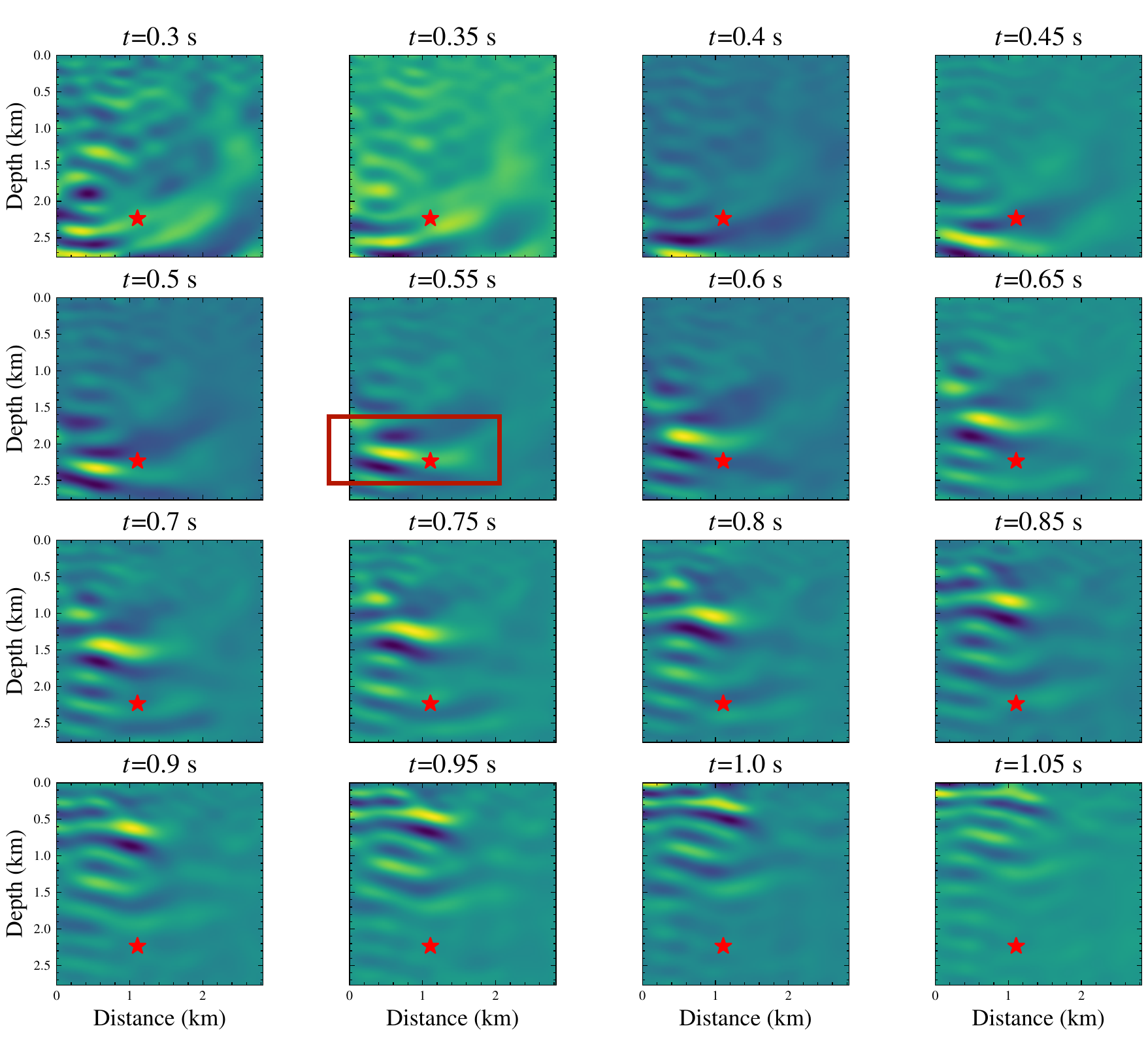}
    \caption{The time-domain snapshots of the wavefield at various times with the proposed method. The red box denotes the location where the energy is focused but is inconsistent with the given label.}
    \label{fig:field_line2_event62_td_image}
\end{figure}

\begin{figure}[!htb]
    \centering
    \includegraphics[width=0.98\columnwidth]{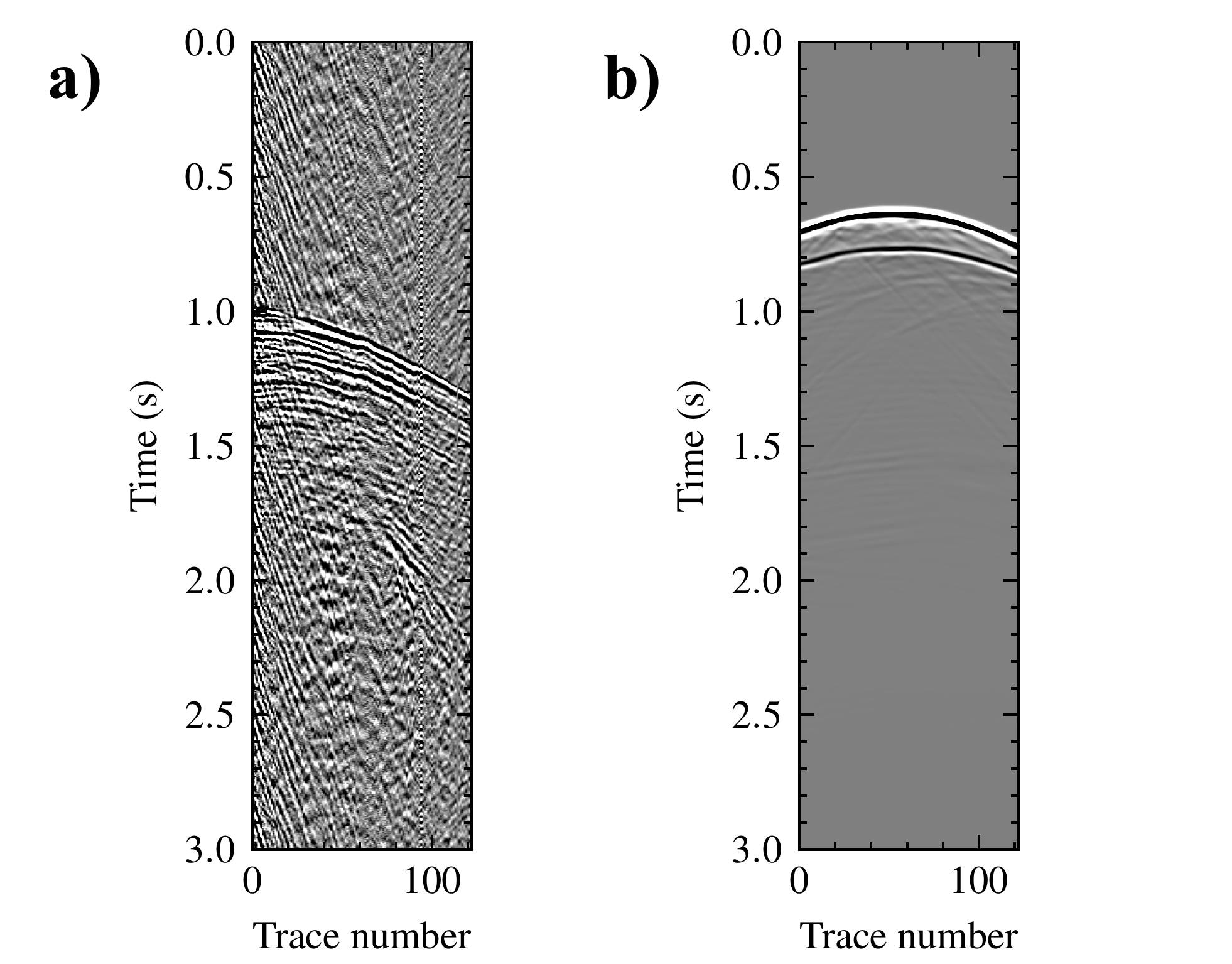}
    \caption{a) is the observed data while b) is the simulated data on the 2D velocity profile (Figure\ref{fig:field_line2_event62_vel_record}b) due to the event on the location given by the data provider.}
    \label{fig:field_line2_event62_sim}
\end{figure}

\begin{figure}[!htb]
    \centering
    \includegraphics[width=1.0\columnwidth]{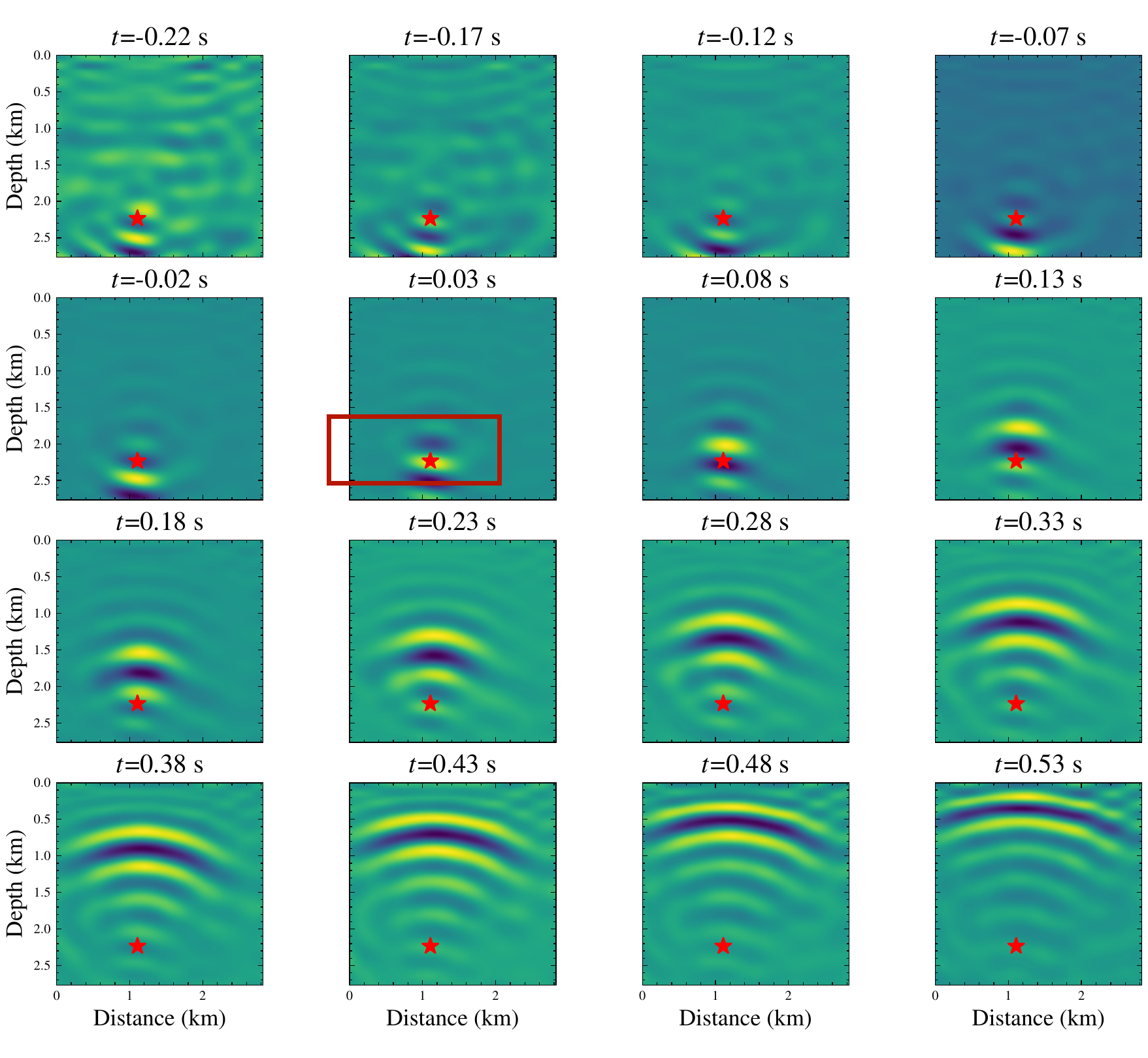}
    \caption{The time domain snapshots of the wavefield at various times with the proposed method applied to the simulated data (Figure\ref{fig:field_line2_event62_sim}b).}
    \label{fig:field_line2_event62_sim_image}
\end{figure}

In the proposed method, we do not need label information for source location during the training as the supervision comes from the governing equations and the data. 
On the other hand, for supervised-based approaches, errors in the labels would result in an error-prone trained model. 
For example, as for event 62 in line 2 (Figures~\ref{fig:field_line2_event62} and~\ref{fig:field_line2_event62_vel_record}), the source image result is shown in Figure~\ref{fig:field_line2_event62_td_image}, where the image focuses at a location different than the provided source location (a potential label for supervised learning). 
We simulate the data from the provided source location label using the 2D P-wave velocity profile and a finite difference wave equation solver. 
We realize that the simulated records and recorded data are very different in curvature and shape (Figure~\ref{fig:field_line2_event62_sim}). 
That means this reference label might be wrong in 2D behavior, while our method may provide the accurate source location at 0.6 s.
To further demonstrate the reliability of the proposed method, using the simulated data, we image the source using our PINN approach and obtain the images in Figure~\ref{fig:field_line2_event62_sim_image}. 
The image focused at the location of the source. 
This demonstrates that the provided source location is not reliable, but our method is consistent, and forms an adjoint to the forward modeling, which is based on a finite difference method.

\section{Discussion}
\label{discussion}
In this paper, we showed the results of our novel 
direct source imaging method using PINNs with hard constraints on both synthetic and field data. 
The functional form of NNs offers flexibility for irregular and sparse recordings. The synthetic demos and field examples demonstrate the effectiveness and potential for source imaging. 
In addition, the proposed method has the potential to be combined with neuron dropout strategies and deep ensemble methods to quantify uncertainty, which we would like to explore in future work.
In the following subsections, we will discuss the causality implementation, the drawbacks of the method, and further potential improvements.

\subsection{The performance with/without causality implementation}
\begin{figure}[!tb]
    \centering
    \includegraphics[width=0.98\columnwidth]{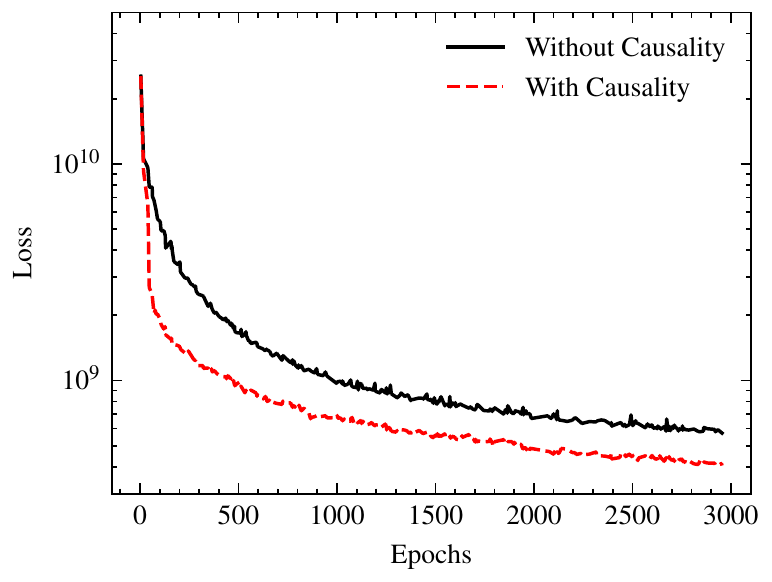}
    \caption{The loss curves of training with and without the causality implementation. The red line is the version without causality implementation, while the black line is with causality implementation.}
    \label{fig:triple-source-loss-cau}
\end{figure}

\begin{figure}[!tb]
    \centering
    \includegraphics[width=0.98\columnwidth]{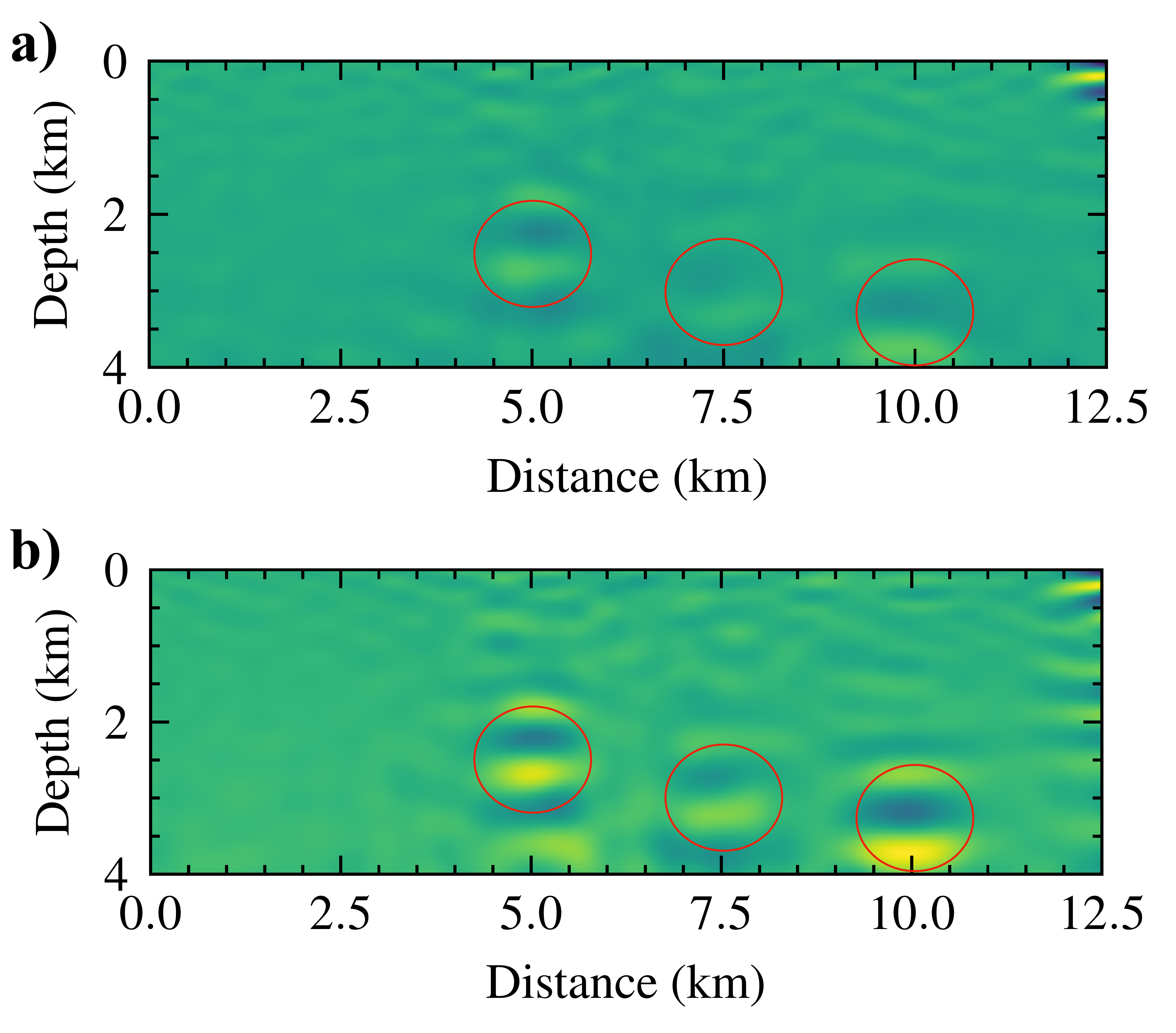}
    \caption{The source image a) with a conventional loss, and b) with a causal loss equation~\ref{equ:causality}.}
    \label{fig:triple-source-cau-com}
\end{figure}

As we claimed earlier, using the PDE loss with causality could accelerate the convergence and improve the final predictions. Here, we take the triple-sources case with 20 random receivers to demonstrate the benefits of the causality implementation. We use the same velocity model to generate the data, the same network configuration but with 3000 epochs in the training, and the same workflow to obtain the source imaging results. For the causality implementation, we set $\epsilon_0=1e-7$ and $\lambda=1e-5$. Figure~\ref{fig:triple-source-loss-cau} shows the loss curves of the training process with and without the causality modification, where the weights were removed from the loss value for a fair comparison. The convergence with causality implementation is much faster than the vanilla implementation. The prediction after 3000 epochs (Figure~\ref{fig:triple-source-cau-com}) with causality implementation is better than the one without causality implementation. The reconstructed source image (Figure~\ref{fig:triple-source-cau-com}b) has better energy focusing.

\subsection{Resolution}
As shown above, the proposed method provides less noisy results compared to the time-reversal methods because the NN provides an automatic anti-alias filter to the data based on the available receiver information. 
However, it is slightly more smeared than the true backpropagation when the receivers are sparse.
Thus, the proposed method still has room for further developments in terms of resolution.
This can be achieved by applying imaging conditions similar to the conventional source imaging methods.
On the other hand, this can also be achieved by extending the frequency range used in the proposed method.
High frequencies have been a thorn for PINNs. However, with the reference frequency approach and the strategy of the frequency extension \cite{huang2022pinnup,huang2022high}, we can gear the approach to work better.
Increasing the frequency range used in the data fitting and PDE fitting would increase the resolution of the final source image.

\subsection{Efficiency}
Although the workflow could deal with irregular and sparse observations and show highly reliable performance, there exists the main drawback of PINNs for now and that is efficiency. The cost is reasonably high, especially considering that the wavefield corresponds to a particular velocity and source location. 
For each new field data measurement, the neural networks need to be retrained. 
However, we can utilize transfer learning \cite{bin2021pinneik}, which can reduce the computational cost.
Overall, PINNs are new compared to, for example, the finite difference methods for solving the wave equation, so speed up is a matter of time. 
Despite its slower implementation, the remarkable flexibility of PINN in its functional wavefield representation, obviating the need for mesh, renders it a compelling avenue for exploration and utilization.

\section{Conclusion}
\label{sec:conclusion}
We present a novel direct microseismic imaging framework utilizing physics-informed neural networks with hard constraints. 
It allows us to image the source due to the use of the neural network functional representation, in addition to its interpolation capabilities. 
Meanwhile, we propose a loss function with causality with respect to depth to accelerate the convergence and improve the prediction of the NN. 
With only 20 random receivers on a 12.5 km lateral stretch, the multiple-source event is stably and accurately located. 
We successfully applied to the Oklahoma Arkoma Basin hydraulic fracturing data, showing the effectiveness of the proposed method and its potential for mesh-free source location. 

\section*{Acknowledgements}
The authors thank KAUST and the DeepWave Consortium sponsors for supporting this research. We thank Microseismic Inc. for the use of the Arkoma data, and Hanchen Wang and Fu Wang for discussing the field data preprocessing. We would also like to thank KAUST for its support and the SWAG group for the collaborative environment. This work utilized the resources of the Supercomputing Laboratory at King Abdullah University of Science and Technology (KAUST) in Thuwal, Saudi Arabia.
\bibliographystyle{IEEEtran}
\bibliography{pinnms}

\ifCLASSOPTIONcaptionsoff
  \newpage
\fi
\end{document}